\documentstyle[aps,epsfig,eqsecnum]{revtex}
\begin{document}
\preprint{}
\draft
\newcommand{\bol}[1]{\mbox{\boldmath $#1$}}
\newcommand{\be}{\begin{eqnarray}}
\newcommand{\ee}{\end{eqnarray}}
\newcommand{\bra}[1]{\mbox{$\langle\, #1 \mid$}}
\newcommand{\bbra}[1]{\mbox{$\left\langle\, #1 \right\mid$}}
\newcommand{\ket}[1]{\mbox{$\mid #1\,\rangle$}}
\newcommand{\bket}[1]{\mbox{$\left\mid #1\,\right\rangle$}}
\newcommand{\pro}[2]{\mbox{$\langle\, #1 \mid #2\,\rangle$}}
\newcommand{\expec}[1]{\mbox{$\langle\, #1\,\rangle$}}
\newcommand{\expecl}[1]{\mbox{$\left\langle\,
            \strut\displaystyle{#1}\,\right\rangle$}}
\newcommand{\real}{\mbox{{\rm I\hspace{-2truemm} R}}}
\renewcommand{\natural}{\mbox{{\rm I\hspace{-2truemm} N}}}
\newcommand{\re}{\real{\rm e}}
\renewcommand{\a}{\hat a}
\newcommand{\ac}{\hat a^{\dagger}}
\renewcommand{\b}{\hat b}
\newcommand{\bc}{\hat b^\dagger}
\title{\bf Non-minimally coupled scalar fields in homogeneous
universes}
\author{G.L. Alberghi\thanks{e-mail: alberghi@bo.infn.it},\,
R. Casadio\thanks{e-mail: casadio@bo.infn.it}\,\, and
A. Gruppuso\thanks{e-mail: gruppuso@bo.infn.it}}
\address{Dipartimento di Fisica, Universit\`a di Bologna, and
I.N.F.N., Sezione di Bologna, via Irnerio 46, 40126 Bologna, Italy}
\maketitle
\begin{abstract}
The equations governing the evolution of non-minimally
coupled scalar matter and the scale factor of a Robertson-Walker
universe are derived from a minisuperspace action.
As for the minimally coupled case, it is shown that the entire
semiclassical dynamics can be retrieved from the Wheeler-DeWitt
equation via the Born-Oppenheimer reduction, which properly yields the
(time-time component of the) covariantly conserved energy-momentum tensor
of the scalar field as the source term for gravity.
However, for a generic coupling, the expectation value of the operator
which evolves the matter state in time is not equal to the source term
in the semiclassical Einstein equation for the scale factor of
the universe and the difference between these two quantities
is related to the squeezing and quantum fluctuations of the matter
state.
We also argue that matter quantum fluctuations become relevant in an
intermediate regime between quantum gravity and semiclassical gravity
and study several cases in detail.
\end{abstract}
\section{Introduction}
Non-minimally coupled scalar fields have been extensively used
in the literature to account for higher loop quantum corrections
to the theory of scalar fields in the general relativity theory
of gravitation (see, {\em e.g.}, \cite{birrell} and Refs. therein).
However, contrary to the minimally coupled case, quantization of
such fields in the framework of quantum field theories on a
classical curved background makes an ambiguity apparent.
In fact, there appear a boundary term in the action for the
scalar field which arises when using rescaled fields and whose
presence changes the canonical matter Hamiltonian \cite{fulling}.
Thus, depending on whether one retains or drops such a term yields
apparently different Schr\"odinger equations for the state of the
scalar field \cite{fgv}.
\par
This ambiguity can be understood by noting that, classically, the
above mentioned boundary term is generated because the rescaling
of the scalar field by a function of the scale factor of the universe
is a time dependent canonical transformation \cite{sud}.
Therefore, in the quantum theory, it is associated with a
(possibly unitary) transformation between different Fock spaces
(choices of the ground state for matter).
The ambiguity can then be eliminated \cite{fgv} by introducing the
invariant Fock space built upon invariant operators \cite{lewis}
which are not affected by a boundary term added to the action.
However, since the matter energy (the time-time component of the
energy-momentum tensor) is related to the matter Hamiltonian,
the interpretation in terms of particle content (or gravitational
``weight'') of a quantum state of matter remains an open question.
\par
The latter is not an academic issue, since, classically, it is
the matter energy which acts as a source for gravity in the
Einstein equations and determines the history of the universe.
Hence, it is of primary concern to clarify by means of which
quantity a quantized state of matter drives the metric at the
semiclassical level.
In order to do so, we appeal to the principle that
time-reparameterization invariance is a fundamental property of
any gravitational system which also generates the dynamics and be
lifted to a quantum symmetry.
Then, one expects to have an Hamiltonian constraint from which the
Born-Oppenheimer (BO) reduction \cite{bv,v,bfv} allows to properly and
unambiguously recover the semiclassical limit starting from the
Wheeler-DeWitt (WDW) equation \cite{dewitt,wheeler}.
\par
The above scheme has proven remarkably successful for the minimally
coupled scalar field in Robertson-Walker space-time \cite{mtw}.
In that case one starts with an action $S=S(N,a,\phi)$ for the
three minisuperspace variables $N$ (lapse function), $a$ (scale
factor) and $\phi$ (one mode of a scalar field) which are functions
of an arbitrary time $t$.
Upon varying $S$ one obtains three Euler-Lagrange equations of motion
(for the details, see next Sections and, {\em e.g.}, Ref.~\cite{bfv})
\be
&&{\delta S\over\delta N}\equiv -H=0\ \ \ \ \ \
\ \ \ \hbox{(Hamiltonian\ constraint)}
\label{H=0}
\\
&&{\delta S\over\delta a}=0
\label{a_eq}
\\
&&{\delta S\over\delta \phi}=0\ \ \ \ \ \
\ \ \ \ \ \
\ \ \ \ \ \ \ \hbox{(Klein-Gordon\ equation)}
\ .
\label{KG}
\ee
Since Eq.~(\ref{H=0}) is a constraint, $N=N(t)$ is arbitrary and one
can set, {\em e.g.}, $N(t)=1$ for $t_i\le t\le t_f$ ({\em proper time}
gauge) provided the initial conditions $a(t_i)$, $\dot a(t_i)$,
$\phi(t_i)$ and $\dot \phi(t_i)$ are such that $H(t_i)=0$ (an overdot
denotes the derivative with respect to $t$).
Then Eqs.~(\ref{a_eq}) and (\ref{KG}) will evolve the initial data
to any time $t_i<t\le t_f$ consistently, since
\be
\left\{\begin{array}{l}
\strut\displaystyle{\delta S\over\delta a}=0 \\
\\
\strut\displaystyle{\delta S\over\delta \phi}=0
\end{array}\right.
\ \ \Rightarrow\ \ {dH\over dt}=0
\ ,
\label{dH}
\ee
which proves that the Hamiltonian constraint is preserved by the dynamics.
However, for practical purposes it is more convenient to revert the above
inference and observe that the equation of motion (\ref{a_eq}) for $a$
is identically satisfied provided the Hamiltonian constraint (\ref{H=0})
is enforced at all times along with the Klein-Gordon equation (\ref{KG}),
\be
\left\{\begin{array}{l}
\strut\displaystyle H(t)=0 \\
\phantom{H(t)=0}\ \ \ (t_i\le t\le t_f) \\
\strut\displaystyle{\delta S\over\delta \phi}=0
\end{array}\right.
\ \ \Rightarrow\ \ {\delta S\over\delta a}=0
\ .
\label{dHH}
\ee
\par
A further simplification is then obtained at the semiclassical level,
where one can show that both the Hamiltonian constraint and
the Klein-Gordon equation follow from the WDW equation obtained by
quantizing the super-Hamiltonian $H$,
\be
\hat H\,\Psi\equiv\left(\hat H_G+\hat H_M\right)\,\Psi=0
\ ,
\label{WdW}
\ee
where $H_G$ and $H_M$ are respectively referred to as the gravitational
and matter Hamiltonian.
In fact, the BO decomposition for the wavefunction of the universe,
\be
\Psi(a,\phi)=\psi(a)\,\chi(\phi,a)
\ ,
\label{BO}
\ee
leads to the coupled equations
\be
&&H_G+\expec{\hat H_M}=0\ \ \ \ \ \ \ \ \
\hbox{(Einstein-Hamilton-Jacobi\ equation)}
\label{HJ}
\\
\nonumber \\
&&i\,\hbar\,{\partial\chi_s\over\partial t}=
\hat H_M\,\chi_s\ \ \ \ \ \ \ \ \
\hbox{(Schr\"odinger\ equation)}
\ ,
\label{schro}
\ee
along with explicit conditions on the wavefunctions $\psi$ and
$\chi$ for the validity of such an approximation.
We then note that Eq.~(\ref{HJ}) is the semiclassical analogue of
Eq.~(\ref{H=0}) and the operator $\hat H_M$ in Eq.~(\ref{schro})
evolves in time states $\chi_s$ of the scalar field in such a way
that the expectation value of $\hat\phi$ over {\em coherent\/} states
satisfies the Klein-Gordon equation (\ref{KG}).
If one regards coherent states as being the quantum states which are
closest to classical, one can conclude that the entire semiclassical
dynamics is encoded in the single WDW equation (\ref{WdW}) and is
properly retrieved by the BO reduction.
\par
We wish to remark here the fact that the decomposition (\ref{BO}) is
not related to the gravitational scale of mass being bigger
than any matter scale, therefore Eqs.~(\ref{HJ}) and
(\ref{schro}) do not require an expansion in powers of the Planck
mass \cite{kiefer} (see also \cite{bfv}, where such an expansion is
shown to violate unitarity when incorrectly performed).
Indeed, the relevant ratio is between the energy of each
quantum of $\phi$ and the total energy of such quanta.
This can be understood if one considers that the scale factor of
the universe is a {\em collective} variable associated with the
total energy of matter in space and that such an energy is presently
much bigger than the energy of each of its microscopic constituents
(described by the degree of freedom $\phi$).
Basically, this fact, the huge amount of matter particles, is the
reason we live in a semiclassical universe \cite{dewitt}.
Consequently, one expects a failure in the semiclassical approximation
which leads to Eqs.~(\ref{HJ}) and (\ref{schro}) for matter states
$\chi$ containing a small number of quanta.
In fact, this expectation has been systematically verified in all
the cases studied so far \cite{cv,cfv,fvv,acvv}.
\par
When one applies the same scheme to the non-minimally coupled case,
a new ambiguity arises because then there is no clear way of splitting
the action into a matter part and a gravitational part and a preferred
classical form for the energy-momentum tensor of the scalar field
can indeed be singled out only by requiring covariant conservation
\cite{madsen}.
Correspondingly, one could write many (classically equivalent)
Hamiltonian constraints which, at the quantum level, become
inequivalent ``WDW'' equations.
Interestingly, the BO factorization (\ref{BO}) pinpoints a specific
form of (or operator ordering in) the WDW equation in order to
ensure the existence of the Schr\"odinger equation (\ref{schro}).
Further, the BO reduction then yields a source term for the geometry
in the semiclassical Einstein equation (\ref{HJ}) which can be easily
related to the time-time component of the proper (divergenceless)
energy-momentum tensor of the scalar field.
However, such a source is not equal to the expectation value of the
Hamiltonian operator in Eq.~(\ref{schro}).
It is only for the case of minimal coupling that the super-Hamiltonian
is the sum of the time-time components of the (covariantly conserved)
energy-momentum tensor and Einstein tensor and then the expectation
value of the generator of the time evolution for the matter state is
the semiclassical source of gravity.
\par
In the following, we shall argue that the difference between the
source term in the semiclassical Einstein equation and the operator
of time-evolution in the Schr\"odinger equation is associated with the
different ways matter and gravity are affected by quantum fluctuations
around the mean value of the matter field and by the presence of a
{\em squeezing} \cite{schumaker} term in the Schr\"odinger equation.
This will lead us to define three different regimes of approximations,
namely {\em quantum gravity}, {\em semiclassical gravity} and an
``intermediate'' regime in which quantum gravitational fluctuations are
negligible but the trajectory of the (collective) gravitational degree
of freedom senses matter fluctuations.
Similar results were found previously in different approaches and,
for example, the intermediate regime is called {\em stochastic gravity}
by Hu and collaborators (for a review see {\em e.g.}, \cite{hu} and
Refs. therein).
\par
The approach employed in the present paper might look totally
different with respect to the procedure of estimating the back-reaction
within the framework of quantum field theory by studying perturbations
of the Einstein equations (or in the Feynman path integral) around a
given classical solution \cite{dewitt1}.
It has certainly the shortcoming that we start from an effective
minisuperspace action in which the degrees of freedom of the system
have been reduced by symmetry arguments prior to quantization and
subsequent semiclassical approximation, rather than from the quantized
set of equations derived from the full Einstein-Hilbert action.
However, we point out that the more standard approach of perturbation
theory also reduces the degrees of freedom of Einstein gravity \cite{hk},
since it assumes the existence of a classical (saddle point) solution
(background manifold and metric) from the onset, leaving as remnant
{\em gauge} freedom only coordinate transformations and small
diffeomorphisms of the background manifold \cite{dewitt1}.
Both approaches are thus questionable if one wishes to quantize
Einstein gravity, but can be regarded as hopefully reliable whenever
one aims at describing gravity in a semiclassical state as we wish to
do here.
\par
In the following Section we start from the minisuperspace action for
a mode of a non-minimally coupled massive scalar field in a
Robertson-Walker space-time and pursue the standard canonical formalism
in order to obtain an Hamiltonian constraint and the corresponding WDW
equation.
Then in Section~\ref{semi} we apply the BO approach and obtain
the Hamilton-Jacobi equation for the scale factor of the universe
and the Schr\"odinger equation for the state of the scalar field.
By making use of invariant operators we solve the Schr\"odinger
equation and show that the source term in the semiclassical Einstein
equation is the semiclassical extension of the time-time component of
the covariantly conserved energy-momentum tensor.
In Section~\ref{special} we specialize the results to the massive
minimally coupled case, which we briefly review, and to the particular
cases with $\xi=1/6$ and $1/4$ which we analyze in more detail.
Finally, in Section~\ref{conc} we summarize and comment on our results.
\par
We follow the sign convention of Ref.~\cite{mtw} and define
$\kappa=8\,\pi\,G$.
\section{Minisuperspace action}
\label{action}
In this Section we shall show that the classical dynamics of a
non-minimally coupled scalar field in Robertson-Walker space-time is
determined by the Hamiltonian constraint and the Klein-Gordon equation,
thus generalizing the result (\ref{dH}) for the minimally coupled case
as described in the Introduction.
\par
The (volume part of the) action for the non-minimally coupled real scalar
field $\Phi=\Phi(x)$ in a generic four-dimensional space-time ${\cal M}$
with metric $\bol g$ is given by \cite{birrell,full}
\be
S_{\Phi}={1\over 2}\,\int_{\cal M} d^4x\,\sqrt{-g}\,\left[
R\,\left({1\over\kappa}-\xi\,\Phi^2\right)
-\left(\partial\Phi\right)^2-\mu^2\,\Phi^2\right]
\ ,
\label{Seh}
\ee
where $g=\det{\bol g}$, $R$ is the scalar curvature, $\mu=1/\ell_\phi$
the inverse of the Compton wavelength of $\Phi$ and $\xi$ a dimensionless
parameter such that $\xi=0$ corresponds to the minimal coupling and
$\xi=1/6$ yields the conformal coupling \cite{miss}.
\par
It is possible to reduce the above action by assuming spatial homogeneity
and isotropy so that ${\cal M}$ admits a preferred foliation into
spatial hypersurfaces of constant time $t$ and the four-metric is given
by the Robertson-Walker line element \cite{mtw}
\be
ds^2=-N^2\,dt^2+a^2\,\left[{dr^2\over 1-k\,r^2}+r^2\,\left(
d\theta^2+\sin^2\theta\,d\varphi^2\right)\right]
\ ,
\label{rw}
\ee
with $k=0,\pm 1$ for flat, positive and negative spatial curvature,
$\theta$ and $\varphi$ the usual angular coordinates and
$r\in[0,r_k)$ with $r_{+1}=1$, $r_0=r_{-1}=+\infty$.
The scalar curvature is then given by
\be
R={6\over N^2}\,\left({\ddot a\over a}-{\dot a\,\dot N\over a\,N}
+{\dot a^2\over a^2}+N^2\,{k\over a^2}\right)
\ .
\label{R}
\ee
We also expand the real scalar field in spatial Fourier modes,
\be
\Phi={1\over\sqrt{2\,V}}\,\sum_{\vec p}\,
\left[e^{-i\,\vec p\cdot\vec x}\,\Phi^{ }_{\vec p}
+e^{i\,\vec p\cdot\vec  x}\,\Phi^\star_{\vec p}\right]
\ ,
\label{sum}
\ee
where ${\cal V}=a^3\,V$ is the spatial volume of the universe,
and separate the real from the imaginary part,
\be
\Phi_{\vec p}(t)={1\over\sqrt{2}}\,
\left(\phi_{\vec p}^1+i\,\phi_{\vec p}^2\right)
\ .
\label{1-2}
\ee
This decomposition yields an effective action for two minisuperspace
variables $a=a(t)$ and $\phi=\phi(t)$ (for the details see
Appendix~\ref{p}),
\be
S={1\over 2}\,\int_{t_i}^{t_f} dt\,\left\{a^3\,\left({\dot\phi^2\over N}
-N\,\omega^2\,\phi^2\right)
+6\,\left(v-\xi\,\phi^2\right)\,
\left[{a\,\dot a^2\over N}+N\,k\,a
+a^2\,{d\over dt}\left({\dot a\over N}\right)\right]
\right\}
\ ,
\label{Sp}
\ee
where $v\equiv V/\kappa$.
The action $S$ can be used to analyze (the real or imaginary part of)
any of the modes $\Phi_{\vec p}$, with an effective frequency $\omega$
as given in Eq.~(\ref{freq}).
\par
The above action contains both a second time derivative of $a$ and a
first time derivative of $N$ in the same term.
The former would cause problems with causality and requires a
modification of the standard Euler-Lagrange equations of motion;
the latter breaks the presumed time-reparameterization invariance
of the system.
However, we observe that upon integrating by parts the last term and
neglecting the integrated part as dynamically irrelevant (for further
explanation see Appendix~\ref{border}), one finally obtains
\be
S={1\over 2}\,\int_{t_i}^{t_f} N\,dt\,a^3\,\left[{\dot\phi^2\over N^2}
-\omega^2\,\phi^2
-{6\over a^2}\,\left(v-\xi\,\phi^2\right)\,
\left({\dot a^2\over N^2}-k\right)
+12\,\xi\,{\phi\over a}\,{\dot a\,\dot\phi\over N^2}\right]
\ ,
\label{S}
\ee
in which there are no second time derivatives and $d\tau\equiv N\,dt$
is the proper time measure.
This is the action we regard as properly describing the dynamics of the
coupled variables $a$ and $\phi$.
\subsection{Lagrangean dynamics}
The Euler-Lagrange equations of motion following from the action $S$
are given by
\be
{1\over a^3}\,{\delta S\over\delta N}&\equiv &-{H\over a^3}
\nonumber \\
&=&
3\,\left(v-\xi\,\phi^2\right)\,
\left({\dot a^2\over a^2}+{k\over a^2}\right)
-{1\over 2}\,\left(\dot\phi^2+\omega^2\,\phi^2\right)
-6\,\xi\,{\dot a\over a}\,\phi\,\dot \phi=0
\label{H1=0}
\\
\nonumber \\
{1\over a^2}\,{\delta S\over\delta a}&=&
3\,\left[
\left(v-\xi\,\phi^2\right)\,\left(2\,{\ddot a\over a}
+{\dot a^2\over a^2}+{k\over a^2}\right)
-4\,\xi\,{\dot a\over a}\,\phi\,\dot \phi
-2\,\xi\,\left(\dot\phi^2+\phi\,\ddot\phi\right)\right.
\nonumber \\
&&\phantom{3\,[}\left.
+{1\over 2}\,\left(\dot\phi^2-\omega^2\,\phi^2\right)
+2\,{\omega\over a^2}\,\vec P\cdot\vec P\,\phi^2\right]=0
\label{a1_eq}
\\
\nonumber \\
{1\over a^3}\,{\delta S\over\delta \phi}&=&
-\left[\ddot\phi+3\,{\dot a\over a}\,\dot\phi+\omega^2\,\phi
+6\,\xi\,\left({\ddot a\over a}+{\dot a^2\over a^2}
+{k\over a^2}\right)\,\phi\right]=0
\label{KG1}
\ ,
\ee
where we have set $N=1$ after the variation to give the expressions
a simple form.
This choice is consistent with the fact that the action (\ref{S})
does not contain time derivatives of $N$ and Eq.~(\ref{H1=0}) is then
the Hamiltonian constraint.
Of course it must be preserved in time and, in fact,
\be
{dH\over d\tau}&=&
-3\,\dot a\,\left(v-\xi\,\phi^2\right)\,
\left(\dot a^2+2\,a\,\ddot a+k\right)
+{3\over 2}\,a^2\,\dot a\,\left(\dot\phi^2+\omega^2\,\phi^2\right)
+a^3\,\dot\phi\,\left(\ddot\phi+\omega^2\,\phi\right)
\nonumber \\
&&
+6\,\xi\,a\,\phi\,\dot\phi\,\left(3\,\dot a^2+k\right)
+6\,\xi\,a^2\,\left(\dot a\,\dot\phi^2+\phi\,\ddot a\,\dot\phi
+\phi\,\dot a\,\ddot\phi\right)
-2\,\omega\,\vec P\cdot\vec P\,\dot a\,\phi^2
\nonumber \\
&=&
-\dot a\,{\delta S\over\delta a}-\dot\phi\,{\delta S\over\delta\phi}
\label{dH1}
\ee
vanishes identically by virtue of Eqs.~(\ref{a1_eq}) and
(\ref{KG1}), thus generalizing to arbitrary $\xi$ the results (\ref{dH})
and (\ref{dHH}) valid for $\xi=0$.
\par
The next step is to quantize the system and show that the WDW equation
encodes the entire semiclassical dynamics.
In order to do so, one needs to consider the Hamiltonian form of
Eq.~(\ref{H1=0}).
\subsection{Hamiltonian dynamics}
\label{hamilton}
The canonical momenta conjugated to $N$, $a$ and $\phi$ are given by
\be
&&P_N=0
\label{PN} \\
\nonumber \\
&&P_a=-6\,a\,\left(v-\xi\,\phi^2\right)\,{\dot a\over N}
+6\,\xi\,a^2\,\phi\,{\dot\phi\over N}
\label{Pa} \\
&& \nonumber \\
&& P_\phi=a^3\,{\dot\phi\over N}+6\,\xi\,a^2\,\phi\,{\dot a\over N}
\ .
\ee
The action (\ref{S}) can then be written in canonical form as
\be
S=\int_{t_i}^{t_f} dt\,\left[P_N\,\dot N+P_a\,\dot a+ P_\phi\,\dot\phi
-N\,H(P_a,P_\phi,a,\phi)\right]
\ ,
\ee
where the super-Hamiltonian already given in Eq.~(\ref{H1=0}) takes on
the rather complicated canonical form
\be
H={1\over 2}\,\left\{
-{\left[a\,P_a-6\,\xi\,\phi\, P_\phi\right]^2
\over 6\,a^3\,\left[v-\xi\,\left(1-6\,\xi\right)\,\phi^2\right]}\,
-6\,k\,a\,\left(v-\xi\,\phi^2\right)
+{ P_\phi^2\over a^3}+a^3\,\omega^2\,\phi^2\right\}
\ .
\label{H}
\ee
\par
Several remarks are in order.
First, the cases $\xi=0$,
\be
H={1\over 2}\,\left\{
-{P_a^2\over 6\,v\,a}-6\,k\,v\,a
+{ P_\phi^2\over a^3}+\,a^3\,\omega^2\,\phi^2\right\}
\ ,
\label{H0}
\ee
and $\xi=1/6$,
\be
H={1\over 2}\,\left\{
-{\left(a\,P_a-\phi\, P_\phi\right)^2\over 6\,v\,a^3}
-6\,k\,v\,a
+{ P_\phi^2\over a^3}+a\,\left(a^2\,\omega^2+k\right)\,\phi^2\right\}
\ ,
\ee
are clearly special since these values of $\xi$ simplify the form of the
kinetic term in $H$, although it is only for $\xi=0$ that the kinetic
term is diagonal in the momenta \cite{diago}.
\par
Second, the $\tau\tau$-component of the unique divergenceless
energy-momentum tensor as computed according to Ref.~\cite{madsen},
\be
T_{\tau\tau}={1\over v-\xi\,\phi^2}\,
\left[{1\over 2}\,\left(\dot\phi^2+\omega^2\,\phi^2\right)
+6\,\xi\,{\dot a\over a}\,\phi\,\dot \phi\right]
\ ,
\label{T}
\ee
is related to the variation of the action with respect to the metric by
\be
{2\over\sqrt{-g}}\,{\delta S_M\over\delta g^{\tau\tau}}=
\left(v-\xi\,\phi^2\right)\,T_{\tau\tau}
\ ,
\label{T_var}
\ee
where
\be
S_M=-{1\over 2}\,\int d^4x\,\sqrt{-g}\,\left[
\left(\partial\Phi\right)^2+\mu^2\,\Phi^2
+\xi\,R\,\Phi^2\right]
\ .
\ee
It is however the (non-conserved) quantity in the right hand side
(r.h.s.) of Eq.~(\ref{T_var}) which naturally appears inside $H$, as is
apparent from Eq.~(\ref{H1=0}) or
\be
{H\over a^3}&=&\left(v-\xi\,\phi^2\right)\,\left[T_{\tau\tau}
-3\,\left({\dot a^2\over a^2}+{k\over a^2}\right)\right]
\nonumber \\
&=&\left(v-\xi\,\phi^2\right)\,\left(T_{\tau\tau}
-{1\over\kappa}\,G_{\tau\tau}\right)
\nonumber \\
&\equiv& \left(v-\xi\,\phi^2\right)\,{H'\over a^3}
\ ,
\label{st_ei}
\ee
where $G_{\tau\tau}$ is the $\tau\tau$-component of the standard Einstein
tensor.
From Eq.~(\ref{dH1}) one concludes that
\be
{dH'\over d\tau}\sim \phi\,\dot\phi\,H'
\ ,
\ee
therefore $G_{\tau\tau}=\kappa\,T_{\tau\tau}$, or $H'=0$,
is an equivalent statement of time-reparameterization invariance
(assuming $v\not=\xi\,\phi^2$).
\par
In order to lift such a symmetry to the quantum level and obtain the WDW
equation one might choose either $H$ or $H'$ (or any other classically
equivalent expression), thus obtaining quantum mechanically inequivalent
``WDW'' equations.
However, the existence of the semiclassical limit via the BO reduction
places some restrictions on the form of the Hamiltonian constraint.
In particular, in order to recover a Schr\"odinger equation from
the WDW equation, it is necessary that the coefficient of $P_a^2$ does
not depend on the matter degree of freedom $\phi$
(see Section~\ref{matter} for more details).
This singles out the preferred (classical) expression
\be
\bar H&=&W_\xi\,H
\nonumber \\
&=&
{1\over 2}\,\left\{
-{\left[a\,P_a-6\,\xi\,\phi\, P_\phi\right]^2\over 6\,a^3}
+W_\xi\,\left[{P_\phi^2\over a^3}+a^3\,\omega^2\,\phi^2
-6\,k\,a\,\left(v-\xi\,\phi^2\right)\,
\right]\right\}
\ ,
\label{bH}
\ee
where $W_\xi=W_\xi(\phi)\equiv v-\xi\,\left(1-6\,\xi\right)\,\phi^2$.
\par
We proceed to analyze the quantum version of the Hamiltonian constraint
with the modified super-Hamiltonian (\ref{bH}) in the next Section where
we shall also show that the BO approach guarantees that the metric be
driven by the proper (conserved) semiclassical ($\tau\tau$-component of
the) energy momentum tensor for the scalar field.
\section{Semiclassical equations}
\label{semi}
The quantization of the Hamiltonian constraint $\bar H=0$, with $\bar H$
as given in Eq.~(\ref{bH}), is formally achieved by introducing the
operators $\hat a$, $\hat \phi$, $\hat P_a$ and $\hat P_\phi$ which
yields the WDW equation
\be
\bar H(\hat P_a,\hat P_\phi,\hat a,\hat\phi)\,\Psi=0
\ ,
\label{WDW}
\ee
where $\Psi=\Psi(a,\phi)$ is the wavefunction of the universe.
\par
When dealing with Eq.~(\ref{WDW}) as it stands, one has to face
several formal problems.
First of all one would like to describe the system by means of Dirac
variables, but, obviously, this is not the case for $a$ and $\phi$ and
their momenta which do not commute with $\bar H$.
This problem could be solved by trading the original canonical variables
for their initial values (the so called {\em perennials}; for a review see
Ref.~\cite{h}).
Then one wishes the canonical variables map into Hermitian operators and,
further, needs to make sense of the ordering in the kinetic term.
\par
Since we are interested in the semiclassical limit for the
variable $a$ \cite{obs}, at each step we shall assume the ordering which
best fits to the computation and define a scalar product in the variable
$\phi$ at fixed $a$ as
\be
\expec{\pro{\Psi}{\Psi'}}(a)=\int_{-\infty}^{+\infty}
d\phi\,\Psi^\star(a,\phi)\,\Psi'(a,\phi)
\ ,
\ee
which renders the operator
\be
\hat P_\phi=-i\,\hbar\,\partial_\phi
\ee
Hermitian provided the functions $\Psi=\Psi(a,\phi)$ are summable in
$\phi\in\real$ for any allowed (fixed) value of $a$.
Analogously we define
\be
\hat P_a=-i\,\hbar\,\partial_a
\ee
and $\hat a$ and $\hat\phi$ as multiplicative operators.
Since the range of $a$ is $\real^+$ one should carefully discuss
the dependence of $\Psi$ on $a$, however, again the fact we want to
recover the semiclassical limit will ease this issue because, strictly
speaking, there is only one allowed value of $a$ at a given time along
a (semi)classical trajectory.
Hence, it will practically be sufficient to consider small intervals
of $\real^+$ (at a time).
\par
To make all the above more concrete, we consider the BO factorization
(\ref{BO}) for the total wavefunction into matter and gravity parts
and assume the matter functions are normalized in the induced scalar
product
\be
\pro{\chi}{\chi'}\equiv
{\expec{\pro{\Psi}{\Psi'}}(a)\over\psi^\star(a)\,\psi'(a)}
=\int_{-\infty}^{+\infty} d\phi\,\chi^\star(\phi,a)\,\chi'(\phi,a)
\ .
\ee
Then we factor out the geometrical phase by defining
\be
&&\tilde\chi=e^{\strut\displaystyle
{-{i\over\hbar}\,\int^a\bra{\chi}\hat P_a\ket{\chi}\,da'}}\,\chi
\nonumber \\
&&\tilde\psi=e^{\strut\displaystyle
{+{i\over\hbar}\,\int^a\bra{\chi}\hat P_a\ket{\chi}\,da'}}\,\psi
\ ,
\ee
so that $\bra{\tilde\chi}\hat P_a\ket{\tilde\chi}=0$.
\subsection{Gravitational equation}
\label{g_semi}
Upon substituting the above definitions into the WDW equation (\ref{WDW})
and contracting over $\bra{\tilde\chi}$ on the left then yields an equation
for the gravitational part
\be
&&{1\over 2}\,\left\{-{\hat P_a^2\over 6\,a}
+2\,\xi\,\expec{\hat\phi\,\hat P_\phi}\,{\hat P_a\over a^2}
-{6\,\xi^2\over a^3}\,\expec{\hat\phi^2\,\hat P_\phi^2}
-6\,k\,a\,\expecl{\hat W_\xi\,\left(v-\xi\,\hat\phi^2\right)}
\right.
\nonumber \\
&&\left.\phantom{{\hat 1\over 2}\,\{}
+{\expec{\hat W_\xi\,\hat P_\phi^2}\over a^3}
+a^3\,\omega^2\,\expec{\hat W_\xi\,\hat\phi^2}\right\}\,\tilde\psi
\nonumber \\
&&
={1\over 2}\,\left\{{\expec{\hat P_a^2}\over 6\,a}\,
-{2\,\xi\over a^2}\,\expec{\hat\phi\,\hat P_\phi\,\hat P_a}
\right\}\,\tilde\psi
\equiv \Delta_a^{(g)}\,\tilde\psi
\ ,
\label{WdWg}
\ee
where $\expec{\hat O}\equiv\bra{\tilde\chi}\hat O\ket{\tilde\chi}$
for any operator $\hat O$ and
$\hat W_\xi=v-\xi\,\left(1-6\,\xi\right)\,\hat\phi^2$.
\par
The term $\Delta_a^{(g)}$ in the r.h.s. of
Eq.~(\ref{WdWg}) can be regarded
as describing fluctuations of the gravitational degree of freedom
in the following sense.
Upon assuming the space of the matter functions $\tilde\chi$ admits
a complete orthonormal basis $\ket{n}$ and inserting an identity
one obtains
\be
\bra{\tilde\chi}f(\hat a,\hat\phi,\hat P_\phi)\,
(\hat P_a)^q\ket{\tilde\chi}&=&
\sum_{n}\,
\bra{\tilde\chi}f(\hat a,\hat\phi,\hat P_\phi)\ket{n}\,
\bra{n}(\hat P_a)^q\ket{\tilde\chi}
\nonumber \\
&\sim&
\sum_{n}\,\pro{n,a+q\,\delta a_n}{\tilde\chi,a}
\ ,
\ee
where $f$ is any function of $\hat a$, $\hat\phi$ and $\hat P_\phi$,
$q$ is a (positive) integer and $\delta a_n\sim
\bra{\tilde\chi}f(\hat a,\hat\phi,\hat P_\phi)\ket{n}$.
Therefore, when $\Delta_a^{(g)}$ is not small [with respect to the
left hand side (l.h.s.)
of Eq.~(\ref{WdWg})] the system has a non-negligible probability of
spreading over states $n(\phi;a')$ with $a'\not=a$.
\par
On the other hand, when $\Delta_a^{(g)}$ is negligible \cite{cl}
one can assume $\tilde\chi$ is peaked on a given trajectory $a=a(\tau)$
and is well approximated by the WKB form
\be
\tilde\psi_{WKB}={1\over\sqrt{-P_a}}\,e^{\strut\displaystyle
{+{i\over\hbar}\,\int P_a(a')\,da'}}
\ ,
\ee
where $P_a=P_a(\tau)$ is the momentum along the semiclassical trajectory
and the time $\tau$ is correspondingly defined according to the
semiclassical version of Eq.~(\ref{Pa}),
\be
P_a=-6\,a\,\expec{\hat W_\xi}\,\dot a
+6\,{\xi\over a}\,\expec{\hat\phi\,\hat P_\phi}
\ .
\label{Pas}
\ee
For
\be
\hat P_a\tilde\psi_{WKB}=P_a\,\tilde\psi_{WKB}
\ ,
\label{WKB}
\ee
one obtains the semiclassical Hamilton-Jacobi equation for $a$,
\be
&&-{P_a^2\over 6\,a}
+2\,\xi\,\expec{\hat\phi\,\hat P_\phi}\,{P_a\over a^2}
-{6\,\xi^2\over a^3}\,\expec{\hat\phi^2\,\hat P_\phi^2}
-6\,k\,a\,\expecl{\hat W_\xi\,\left(v-\xi\,\hat\phi^2\right)}
\nonumber \\
&&
+{\expec{\hat W_\xi\,\hat P_\phi^2}\over a^3}
+a^3\,\omega^2\,\expec{\hat W_\xi\,\hat\phi^2}=0
\ ,
\ee
or, after substituting for $P_a$ from Eq.~(\ref{Pas}),
\be
&&
3\,a\,\left[\dot a^2
+k\,{\expecl{\hat W_\xi\,\left(v-\xi\,\hat\phi^2\right)}\over
\expec{\hat W_\xi}^2}
\right]
-{1\over 2}\,\left[
{\expec{\hat W_\xi\,\hat P_\phi^2}\over\expec{\hat W_\xi}^2\,a^3}
+a^3\,\omega^2\,
{\expec{\hat W_\xi\,\hat\phi^2}\over\expec{\hat W_\xi}^2}\right]
\nonumber \\
&&=-3\,\xi^2\,{\expec{\hat\phi^2\,\hat P_\phi^2}
-\expec{\hat\phi\,\hat P_\phi}^2\over\expec{\hat W_\xi}^2\,a^3}
\equiv \Delta_\phi
\ .
\label{g}
\ee
This expression shows a deep entanglement between the two degrees of
freedom of the system, so that it is not possible to distinguish
a gravitational Hamiltonian from a matter Hamiltonian uniquely, as
was done in Eq.~(\ref{HJ}), when $\xi\not=0$.
\par
The meaning of $\Delta_\phi$ in the r.h.s.~of Eq.~(\ref{g}) is that
it describes quantum fluctuations of the matter field.
In fact, such a term corresponds to higher $\hbar$-order terms in
the expansion of $\tilde\chi$ around a classical state,
for which
\be
\expec{\hat\phi^2\,\hat P_\phi^2}=\expec{\hat\phi\,\hat P_\phi}^2
+{\cal O}(\hbar)
\ ,
\ee
thus $\Delta_\phi$ measures departures from classicality of the scalar
field and could also be associated with fluctuations of the ``effective''
Newton constant, which would explain why $\Delta_\phi\sim\xi^2$
\cite{frolov}.
These fluctuations are formally different from the ones described
by $\Delta_a^{(g)}$ and associated with the classicality of the
gravitational degree of freedom which appear in Eq.~(\ref{WdWg}).
However, it is also expected that both kinds of fluctuations must be
consistently small in a semiclassical regime.
In fact, terms like $\Delta_\phi$ are usually absent in the standard
expressions for the source in the semiclassical Einstein equations as
obtained in quantum field theory in Robertson-Walker space
\cite{birrell,hu}.
\par
When $\Delta_\phi$ is negligible, Eq.~(\ref{g}) simplifies to
\be
3\,a\,\left[\expec{\hat W_\xi}\,\dot a^2
+k\,\left(v-\xi\,\expec{\hat\phi^2}\right)\right]
\simeq{1\over 2}\,\left[{\expec{\hat P_\phi^2}\over a^3}
+a^3\,\omega^2\,\expec{\hat\phi^2}\right]
\ ,
\label{sg}
\ee
which then becomes equal to the classical Hamiltonian constraint
(\ref{H1=0}) if we replace $\expec{\hat\phi^2}$ with $\phi^2$ and
$\expec{\hat P_\phi^2}$ with $P_\phi^2$.
This shows that the BO approach yields the correct classical limit
for gravity.
\par
At this point it might help in further understanding the three equations
(\ref{WdWg}), (\ref{g}) and (\ref{sg}) derived so far to recall the three
regimes of approximation which were mentioned in the Introduction and give
for them explicit definitions in terms of the relevant quantities:
\begin{enumerate}
\item
\label{quantumg}
when $\Delta_a^{(g)}$ and $\Delta_\phi$ are not negligible [with respect to
the corresponding l.h.s.s in Eqs.~(\ref{WdWg}) and (\ref{g})],
one is at the border of {\em quantum gravity} in which gravitational
fluctuations start to be significant.
All the present treatment of the coupled matter-gravity system, and
including the starting action (\ref{S}), is likely to lose its meaning
when $\Delta_a^{(g)}$ and $\Delta_\phi$ are very big, but one can still try
to use Eq.~(\ref{WdWg}) whenever such terms are not overwhelming;
\item
\label{semig}
when both $\Delta_a^{(g)}$ and $\Delta_\phi$ are negligible, one is at
the opposite limit of {\em semiclassical gravity} described by
Eq.~(\ref{sg}).
This provides a good picture of present day universe and already
contains the description of important effects related to the quantum
nature of matter such as the production of particles induced by the
evolution of the scale factor $a=a(t)$ \cite{birrell,fvv};
\item
\label{interg}
when $\Delta_a^{(g)}$ is negligible, but $\Delta_\phi$ is not, one is
in the ``intermediate'' regime where matter fluctuations play a
significant role in driving the evolution of the metric according to
Eq.~(\ref{g}).
One might expect that this is a fairly good setting for the description
of an early epoch in the history of the universe, {\em e.g.}, near
the time of the onset of inflation \cite{hu}.
\end{enumerate}
We also point out that, in the above scheme, one could include within
matter fluctuations the effect of higher WKB orders in the expression
for $\tilde\psi$ (representing ``collective gravitational fluctuations''),
which might play a significant role at both stages~\ref{semig} and
\ref{interg} \cite{bob}.
\par
We finally note that, so far, we have not chosen a specific ordering
between $\hat\phi$ and $\hat P_\phi$ to keep the discussion as
general as possible.
\subsection{Matter equation}
\label{matter}
Let us now turn to the equation for the matter state.
Upon subtracting Eq.~(\ref{WdWg}) from Eq.~(\ref{WDW}) and dividing by
$\tilde\psi=\tilde\psi_{WKB}$ one obtains
\be
&&\left\{{P_a\over 2}\,\left[-{\hat P_a\over 3\,a}
-2\,{\xi\over a^2}\,\expec{\hat\phi\,\hat P_\phi}
+2\,{\xi\over a^2}\,\hat\phi\,\hat P_\phi\right]
+\hat H_S'-\expec{\hat H_S'}\right\}\,\tilde\chi
\nonumber \\
&&={1\over 2}\,\left[{\hat P_a^2-\expec{\hat P_a^2}\over 3\,a}
-2\,{\xi\over a^2}\,\left(\hat\phi\,\hat P_\phi\,\hat P_a
-\expec{\hat\phi\,\hat P_\phi\,\hat P_a}\right)\right]\,\tilde\chi
\ ,
\label{WdWm}
\ee
where we have also used Eq.~(\ref{WKB}) and
\be
\hat H_S'\equiv
{\hat W_\xi\over 2}\,
\left[{\hat P_\phi^2\over a^3}+a^3\,\omega^2\,\hat\phi^2
-6\,k\,a\,\left(v-\xi\,\hat\phi^2\right)\right]
-3\,\xi^2\,{\hat\phi^2\,\hat P_\phi^2\over a^3}+3\,k\,a\,v^2
\ .
\ee
\par
From Eq.~(\ref{Pas}) one obtains
\be
P_a\,\hat P_a\sim i\,\hbar\,\dot a\,{\partial\over\partial a}
\equiv i\,\hbar\,{\partial\ \over\partial \tau}
\ ,
\ee
and this explains our previous statement that one cannot allow
for a factor $f(\phi)$ multiplying $P_a^2$ in the super-Hamiltonian.
In that case one would have
\be
f(\hat\phi)\,P_a\,\hat P_a\tilde\chi
\sim f(\hat\phi)\,{\partial\tilde\chi\over\partial\tau}
\ .
\ee
Then, the only way of getting rid of the operator $f(\hat\phi)$
and obtain a Schr\"odinger equation is to assume the ordering
\be
\left[f(\hat\phi)\,{\partial\over\partial\tau}\right]\,\tilde\chi
\equiv{\partial\over\partial\tau}\,\left[f(\hat\phi)\,
\tilde\chi\right]
\ ,
\ee
and write
\be
\tilde\chi={1\over f(\phi)}\,\bar\chi
\ .
\ee
However, one would eventually find that this procedure is equivalent
to the (simpler) choice of multiplying $H$ by $1/f(\phi)$ from the onset,
as was indeed done by introducing $\bar H$ in Eq.~(\ref{bH}).
\par
Upon substituting in for $P_a$ from Eq.~(\ref{Pas}) yields
\be
i\,\hbar\,{\partial\tilde\chi\over\partial\tau}=
\left[\hat H_S-\expec{\hat H_S}\right]\,\tilde\chi
+\Delta_a^{(m)}\,\tilde\chi
\ ,
\ee
where
\be
\hat H_S&\equiv&{1\over\expec{\hat W_\xi}}\,\left\{
{\hat W_\xi\over 2}\,
\left[{\hat P_\phi^2\over a^3}+a^3\,\omega^2\,\hat\phi^2
-6\,k\,a\,\left(v-\xi\,\hat\phi^2\right)\right]
+3\,k\,a\,v^2\right\}
-6\,\xi\,{\dot a\over a}\,\hat\phi\,\hat P_\phi
\nonumber \\
&&+3\,\xi^2\,{2\,\expec{\hat\phi\,\hat P_\phi}\,\hat\phi\,\hat P_\phi
-\hat\phi^2\,\hat P_\phi^2\over\expec{\hat W_\xi}\,a^3}
\ ,
\label{HS}
\ee
and
\be
\Delta_a^{(m)}\equiv
{1\over\expec{\hat W_\xi}}\,
\left[{\xi\over a^2}\,\left(\hat\phi\,\hat P_\phi\,\hat P_a
+\expec{\hat\phi\,\hat P_\phi}\,\hat P_a
-\expec{\hat\phi\,\hat P_\phi\,\hat P_a}\right)
-{\hat P_a^2-\expec{\hat P_a^2}\over 12\,a}
\right]
\ee
represents the effect of gravitational fluctuations on the evolution
of matter as $\Delta_a^{(g)}$ does for the gravitational state.
\par
Finally, by neglecting $\Delta_a^{(m)}$ and rescaling the matter
wavefunction,
\be
\chi_s\equiv e^{\strut\displaystyle
{-{i\over\hbar}\,\int d\tau\,\expec{\hat H_S}}}\,\tilde\chi
=e^{\strut\displaystyle
{-{i\over\hbar}\,\int d\tau\,\expec{\hat H_S-i\,\hbar\,\partial_\tau}}}\,
\chi
\ ,
\label{2phases}
\ee
one obtains the Schr\"odinger equation
\be
i\,\hbar\,{\partial\chi_s\over\partial\tau}=\hat H_S\,\chi_s
\ .
\label{sc}
\ee
\par
One can make use of Eq.~(\ref{g}) to substitute for $\dot a$ in
the operator $\hat H_S$ which appears in the r.h.s. above and
generates the evolution of the matter state along the semiclassical
trajectory $a=a(\tau)$, thus obtaining a fairly complicated
expression which does not naively compare to the matter source that
appears in the semiclassical Einstein equation (\ref{g}).
\par
For a generic value of $\xi$ it looks hopeless to find {\em invariant
operators} \cite{lewis} for the complicated Hamiltonian $H_S$, but it
can be done at least when terms of order $v^{-1}$ are negligible
(along with $\Delta_a^{(m)}\sim\Delta_a^{(g)}$; see case~\ref{semig}
in the scheme of approximations of Section~\ref{g_semi}).
Then
\be
\hat W_\xi\sim v
\label{Wv}
\ ,
\ee
and Eq.~(\ref{HS}) reduces to
\be
\hat H_S\simeq
{1\over 2}\,\left[{\hat P_\phi^2\over a^3}
+a\,\left(a^2\,\omega^2+6\,\xi\,k\right)\,\hat\phi^2
-6\,\xi\,{\dot a\over a}\,\left(\hat\phi\,\hat P_\phi+
\hat P_\phi\,\hat\phi\right)\right]
\ ,
\label{Hs}
\ee
where we have also symmetrized the product $\hat\phi\,\hat P_\phi$.
Such a term is related to the {\em squeezing} of the matter state
(see, {\em e.g.} the review \cite{schumaker}) and it is remarkable that
it appears in the operator that evolves the state of the field $\phi$.
In fact, upon quantizing the scalar field in a Robertson-Walker
universe, one obtains a squeezing term in the Hamiltonian for the
rescaled field $\zeta\equiv a\,\phi$, which is however absent in the
Hamiltonian for the unscaled $\phi$ \cite{fgv}.
Since the squeezing is also related to the decoherence of classical
solutions \cite{kief,hu}, this might have interesting consequences for
the onset of a classical universe.
\par
It is important to note that neglecting terms of order $v^{-1}=\kappa/V$
is tantamount to perform an expansion in the Newton constant and
neglect terms of order $\kappa$ and higher.
In fact, a (infinite) factor of $V$ has been absorbed in the definition
of $\phi^2$ [see Eq.~(\ref{sum})], therefore one can formally set $V=1$
and take $v=\kappa^{-1}$ henceforth.
As explained in Ref.~\cite{bfv}, the expansion in $\kappa$ must be
performed carefully in order to preserve unitarity and, indeed, in
the present paper we expand after completing the BO reduction which
yields the Eqs.~(\ref{g}) and (\ref{HS}).
Had we expanded and truncated the original super-Hamiltonian (\ref{H1=0})
before applying the BO approach might have led us to an unphysical picture
in which, {\em e.g.}, one neglects the matter source (of order $\kappa^0$)
with respect to the gravitational part (of order $\kappa^{-1}$).
\par
For the above approximate Hamiltonian $\hat H_S$ one finds that the
annihilation and creation operators $\a_\xi$ and $\ac_\xi$ are given by
\be
\a_\xi=\sqrt{a^3\,\omega_\xi\over 2\,\hbar}\,
\left[\hat\phi+{i\over\omega_\xi}\,\left(
{\hat P_\phi\over a^3}-6\,\xi\,{\dot a \over a}\,\hat\phi\right)
\right]
\ ,
\label{a}
\ee
where the effective frequency is
\be
\omega_\xi^2\equiv\omega^2+6\,\xi\,\left({k\over a^2}
-6\,\xi\,{\dot a^2\over a^2}\right)
\ .
\ee
The latter quantity must be strictly positive, that is \cite{gao}
\be
36\,\xi^2\,{\dot a^2\over a^2}<
\omega^2+6\,\xi\,{k\over a^2}
\ ,
\label{cond}
\ee
which is always satisfied for a minimally coupled massive ($\mu\not=0$)
scalar field, but places restrictions on the (Hubble) ratio $h=\dot a/a$
for all other values of the parameters $\mu$, $\vec P\cdot\vec P$ and $k$.
In particular, the case $\omega=k=0$ is excluded from the present
analysis.
The Hamiltonian can then be written as
\be
\hat H_S=\hbar\,\omega_\xi\,\left(\ac_\xi\,\a_\xi+{1\over 2}\right)
\ ,
\ee
however, this does not imply that the model contains states
with a fixed number of quanta.
\par
In fact, exact solutions of Eq.~(\ref{sc}) are constructed from the
invariant operators $\b_\xi$ and $\bc_\xi$ \cite{gao},
\be
\b_\xi={1\over \sqrt{2\,\hbar}}\,\left\{{\hat\phi\over\rho_\xi}
+i\,\left[\rho_\xi\,\hat P_\phi
-\left(a^3\,\dot\rho_\xi+6\,\xi\,a^2\,\dot a\,\rho_\xi\right)\,\hat\phi
\right]\right\}
\ ,
\ee
according to
\be
\ket{n,\tau}_s=e^{\strut\displaystyle
{-{i\over\hbar}\,\int^\tau d\tau'\,_s\bra{n,\tau'}\,
\hat H_S-i\,\hbar\,\partial_\tau\,\ket{n,\tau'}_s}}\,
\ket{n}_\xi
\ ,
\label{chi_s}
\ee
where
\be
\ket{n}_\xi\equiv {\left(\bc_\xi\right)^n\over\sqrt{n!}}\,\ket{0}_\xi
\ ,
\ee
and we remark that the phase in Eq.~(\ref{chi_s}) is the same that
relates $\chi_s$ to (the time-independent) $\chi$ in Eq.~(\ref{2phases}).
The function $\rho_\xi=\rho_\xi(\tau)$ is a solution of the equation
\be
\ddot\rho_\xi+3\,{\dot a\over a}\,\dot\rho_\xi
+\left[\omega^2+\xi\,\left(R+6\,\left(1-6\,\xi\right)\,{\dot a^2\over a^2}
\right)\right]\,\rho_\xi
={1\over a^6\,\rho_\xi^3}
\ ,
\label{rho}
\ee
with $R$ the scalar curvature (\ref{R}) in the gauge $N=1$.
We observe that setting the r.h.s. of the above equation to zero yields
the classical Klein-Gordon equation (\ref{KG1}) only for the two cases
$\xi=0,1/6$.
\par
By introducing the new variable
\be
\sigma_\xi\equiv a^{3/2}\,\rho_\xi
\ ,
\ee
one simplifies Eq.~(\ref{rho}) to the form
\be
\ddot\sigma_\xi+\Omega^2_\xi\,\sigma_\xi={1\over\sigma_\xi^3}
\ ,
\label{pin}
\ee
where
\be
\Omega^2_\xi=\omega^2+6\,\xi\,{k\over a^2}
+{3\over 4}\,\left[16\,\xi\,\left(1-3\,\xi\right)-1\right]
\,{\dot a^2\over a^2}
+{3\over 2}\,\left(4\,\xi-1\right)\,{\ddot a\over a}
\ .
\ee
We then note that another special case is obtained for $\xi=1/4$ in
flat space, since $\Omega_{1/4}=\omega$ for $k=0$.
This case is of interest also because every solution $\eta$ of the (free)
Dirac equation in curved space-time satisfies the Klein-Gordon equation
\be
\left(\Box+\mu^2+\xi_f\,R\right)\eta=0
\ ,
\ee
where $\xi_f=1/4$ \cite{birrell,casta}.
Therefore, one can consider the scalar field with $\xi=1/4$ as the
analogue of a minimally coupled fermion field.
\par
Provided $\ddot\sigma_\xi$ is negligible, a solution to Eq.~(\ref{pin}) is
given by
\be
\sigma_\xi&\simeq&\Omega_\xi^{-1/2}
\ .
\label{expan}
\ee
One then finds that $\,_s\bra{n,\tau}\,\hat N\,\ket{n,\tau}_s
\equiv\,_s\bra{n,\tau}\,\ac_\xi\,\a_\xi\,\ket{n,\tau}_s$ is generally
a time-dependent quantity.
Also, in general there exist suitable initial conditions (at $\tau=\tau_i$)
such that
\be
\left\{\begin{array}{l}
\sigma_\xi(\tau_0)=
\strut\displaystyle{1\over\sqrt{\omega_\xi(\tau_0)}}
\\
\\
\dot\sigma_\xi(\tau_0)=\strut\displaystyle{3\dot a(\tau_0)\over2\,a(\tau_0)}\,
\sigma_\xi(\tau_0)
\ ,
\end{array}\right.
\ \ \ \Rightarrow\ \ \
\b_\xi(\tau_0)=\a_\xi(\tau_0)
\ ,
\label{in}
\ee
and the state $\ket{n,\tau}_s$ then describes exactly $n$ (Hamiltonian)
quanta at $\tau=\tau_0$.
In the following we shall often find it convenient to consider
$\tau_0\gg\tau_i$ corresponding to a period of (almost) adiabatic
expansion with $\dot a\ll 1$.
\par
We conclude this part by recalling that it is usually stated that the
correct way of counting particles in quantum field theory on curved
backgrounds is not by means of the number operator $\hat N\sim\hat H_S$
but by introducing a (localized) detector coupled to the matter field
\cite{birrell}.
In the present context, one might also dispute that there is the
alternative option of counting the number of particles by making use of
their ``weight'' as it appears in Eq.~(\ref{g}), since this is what
drives the observable $a$.
We shall further investigate this point in the next Sections.
\subsection{Coupled dynamics}
\label{couple}
In the same (small $\kappa$) approximation (\ref{Wv}) which led us to the
Schr\"odinger operator $\hat H_S$ in Eq.~(\ref{Hs}), the gravitational
equation (\ref{sg}) becomes
\be
3\,\left({\dot a^2\over a^2}+{k\over a^2}\right)=
\kappa\,{\expec{\hat H_M}\over a^3}
\ ,
\label{sgv}
\ee
where we have neglected $\Delta_\phi\sim v^{-2}$ and ($v\sim\kappa^{-1}$)
\be
\hat H_M={1\over 2}\,\left[{\hat P_\phi^2\over a^3}
+a\,\left(a^2\,\omega^2+6\,\xi\,k\right)\,\hat\phi^2\right]
\ .
\ee
It then follows that the expectation value
$H_s\equiv\,_s\bra{n,\tau}\hat H_S\,\ket{n,\tau}_s$ can be expressed
in terms of the function $\sigma_\xi$ as \cite{gao,super}
\be
H_s={\hbar\over 2}\,\left(n+{1\over 2}\right)\,
\left\{\left[\omega^2+6\,\xi\,{k\over a^2}
+\left({9\over 4}-36\,\xi^2\right)\,{\dot a^2\over a^2}\right]\,
\sigma_\xi^2
+\dot\sigma_\xi^2
-3\,{\dot a\over a}\,\sigma_\xi\,\dot\sigma_\xi
+{1\over\sigma_\xi^2}\right\}
\ ,
\label{Hsn}
\ee
and, after subtracting the squeezing term, one finally obtains the
``weight'' $H_m\equiv\,_s\bra{n,\tau}\hat H_M\,\ket{n,\tau}_s$
as
\be
H_m={\hbar\over 2}\,\left(n+{1\over 2}\right)\,
\left\{\left[\omega^2+6\,\xi\,{k\over a^2}
+\left({9\over 4}+18\,\xi\,\left(2\,\xi-1\right)\right)\,
{\dot a^2\over a^2}\right]\,\sigma_\xi^2
+\dot\sigma_\xi^2
+3\,\left(4\,\xi-1\right)\,{\dot a\over a}\,\sigma_\xi\,\dot\sigma_\xi
+{1\over\sigma_\xi^2}\right\}
\ .
\label{Hmn}
\ee
\par
At least for $a$ large, one expects that terms proportional to $\dot a$
and higher time derivatives of $a$ become small [with respect to $\omega$ or
$k/a$, see condition (\ref{cond})].
In this approximation one then has, to next to leading order,
\be
\begin{array}{l}
\sigma_\xi\simeq\strut\displaystyle
\left(\omega^2+6\,\xi\,{k\over a^2}\right)^{-1/4}\,
\left(1+\strut\displaystyle{3\over 16}\,
{1-16\,\xi\,\left(1-3\,\xi\right)\over
\omega^2+6\,\xi\,{k\over a^2}}\,{\dot a^2\over a^2}
+\strut\displaystyle{3\over 8}\,
{1-4\,\xi\over\omega^2+6\,\xi\,{k\over a^2}}\,
{\ddot a\over a}\right)
\\
\\
\dot\sigma_\xi\simeq\strut\displaystyle{\dot a\over a}\,
\strut\displaystyle\left(\omega^2+6\,\xi\,{k\over a^2}\right)^{-5/4}\,
\left({\vec P\cdot\vec P\over2\,a^2}+3\,\xi\,{k\over a^2}\right)
\ .
\end{array}
\label{expan2}
\ee
The next step is to substitute $H_m$ in Eq.~(\ref{Hmn}) with the above
expression for $\sigma_\xi$ into Eq.~(\ref{sgv}) which yields a {\em master
equation} for the scale factor of the universe $a=a(\tau)$.
\par
This master equation can then be integrated (at least numerically) for
different choices of the parameters.
The latter must be so chosen that the following conditions hold:
\begin{enumerate}
\item\label{(i)}
$(\dot a/a)^2$ and $\ddot a/a$ must be small with respect to
$\omega^2+k/a^2$, as is required by the condition (\ref{cond}), and the
approximate expression (\ref{expan2}) can then be employed;
\item\label{(ii)}
the number of invariant quanta $n\gg 1$ in order for $\Delta_a^{(g)}$ and
$\Delta_a^{(m)}$ to be negligible \cite{cv,cfv,fvv,acvv};
\item\label{(iii)}
the effective volume of the universe $v\,a^3/n\gg 1/\omega$ so that
$\Delta_\phi$ is negligible and $\hat H_S$ can be approximated by
Eq.~(\ref{Hs}).
This last condition can also be written as
\be
n\ll {a^3\over\ell_p^2\,\ell_\phi}
\ ,
\label{dilute}
\ee
which is like a dilute gas approximation, and
$\ell_p\equiv\sqrt{\hbar\,\kappa}$ is the Planck length.
\end{enumerate}
In the following Section we shall try and relax the condition~\ref{(iii)}
to estimate the effects of matter fluctuations.
\par
As we mentioned at the end of Section~\ref{matter}, one could think of using
the quantity $H_m$ to measure the actual energy of matter in the universe.
Thus, one is led to identify $a^3\,\expec{\hat T^\tau_{\ \tau}}=
{\cal V}\,\expec{\hat T^\tau_{\ \tau}}=-H_m$
and impose the conservation law
\be
{\bol \nabla}\cdot\expec{\hat{\bol T}}=0
\ ,
\label{cons}
\ee
which is consistent with the fact that the l.h.s. of Eq.~(\ref{sgv}) is
recognized as the ($\tau\tau$-component of the) standard Einstein tensor
for the Robertson-Walker metric [see Eq.~(\ref{st_ei})].
As anticipated at the end of Section~\ref{hamilton}, this shows that the
BO reduction automatically provides the correct expression for the
energy-momentum tensor in the semiclassical Einstein equations.
\par
Since the scalar field is (classically) equivalent to a perfect fluid
\cite{mtw,madsen}, we can assume the preferred foliation of the space-time
${\cal M}$ in which the four-metric takes the form (\ref{rw})
corresponds to the frame comoving with this fluid of (gravitating) energy
$H_m$ (see note~\cite{obs}).
The (semiclassical) energy-momentum tensor of the scalar field in the
comoving frame can be written as \cite{mtw}
\be
\expec{\hat{\bol T}}={\rm diag\,}\left[-{H_m\over{\cal V}},p,p,p\right]
\ ,
\label{Tf}
\ee
where $p$ is the pressure.
The spatial components of Eq.~(\ref{cons}) then imply that $p=p(\tau)$, in
agreement with the hypothesis of homogeneity and isotropy, and the
$\tau$-component of Eq.~(\ref{cons}) yields the expected relation
\be
p=-{\delta H_m\over\delta{\cal V}}
=-{\dot H_m\over3\,\dot a\,a^2}
\ ,
\ee
which can be used to determine the pressure once $H_m$ and $a$ have been
obtained.
Here we only observe that one has dust ($p=0$) whenever $H_m$ is constant.
\section{Special cases}
\label{special}
It is clear from the previous analysis that the cases $\xi=0,1/6$
are particularly simple for a variety of reasons, including the fact
that $W_0=W_{1/6}=v$.
Thus, we now review the minimally coupled case and study in detail the
conformally coupled case.
We shall also consider the case with $\xi=1/4$ (the analogue of a minimally
coupled fermion field) for which the invariant structure is particularly
simple in the limit (\ref{Wv}) when $\vec P=k=0$.
\subsection{Minimal coupling}
For $\xi=0$ the separation between matter and gravity is clear from
the outset, since the semiclassical equation (\ref{sg}) for $a$ is given
by
\be
3\,a\,\left(\dot a^2+k\right)
={\kappa\over 2}\,\left({\expec{\hat P_\phi^2}\over a^3}
+a^3\,\omega^2\,\expec{\hat\phi^2}\right)
\ ,
\label{g0}
\ee
where the $\tau\tau$-component of the energy-momentum tensor,
\be
a^3\,\expec{\hat T_{\tau\tau}}={1\over 2}\,
\left({\expec{\hat P_\phi^2}\over a^3}
+a^3\,\omega^2\,\expec{\hat\phi^2}\right)
\ ,
\ee
equals the expression that is obtained by quantizing the field $\phi$ on
the Robertson-Walker background and taking for the energy density the
expression given in Eq.~(\ref{T}) for $\xi=0$.
This shows the equivalence of the BO approach for the minimally
coupled scalar field to the computations performed in the more
common framework of quantum field theory on curved backgrounds.
\par
One also has the identity $\kappa\,\hat H_S=\hat H_M$ and the invariant
annihilation operator, for the cases when $\omega\not=0$
[see Eq.~(\ref{cond})], reduces to the simple form
\be
\b_0={1\over \sqrt{2\,\hbar}}\,\left[{\hat\phi\over\rho_0}
+i\,\left(\rho_0\,\hat P_\phi-a^3\,\dot\rho_0\,\hat\phi\right)\right]
\ ,
\ee
with the function $\rho_0=\rho_0(\tau)$ determined by the equation
\be
\ddot\rho_0+3\,{\dot a\over a}\,\dot\rho_0+\omega^2\,\rho_0
={1\over a^6\,\rho_0^3}
\ .
\ee
This yields the exact (invariant) Fock space of states $\ket{n}_0$
with which one can also build coherent states,
\be
\b_0\,\ket{\alpha}_0=\alpha\,\ket{\alpha}_0
\ ,
\ee
such that the expectation value
\be
\phi_c(\tau)\equiv\,_s\bra{\alpha,\tau}\,\hat\phi\,\ket{\alpha,\tau}_s
\ ,
\ee
satisfies the classical Klein-Gordon equation (\ref{KG1}) for $\xi=0$
\cite{fvv}
\be
\ddot\phi_c+3\,{\dot a\over a}\,\dot\phi_c+\omega^2\,\phi_c=0
\ .
\ee
This is the last step required to show that the semiclassical dynamics
of $a$ and $\phi$ can be retrieved from the WDW equation alone when
the scalar field is minimally coupled.
\par
For the homogeneous mode in flat space, $\vec P=k=0$, the approximation
(\ref{expan2}) yields
\be
H_m=H_s\simeq m_\phi\,\left(n+{1\over 2}\right)\,
\left[1+{9\,\ell_\phi^2\,\dot a^2\over8\,a^2}\right]
\ ,
\ee
where $m_\phi=\hbar\,\mu$ is the (inertial) mass of one scalar quantum.
A part from the value of the numerical factor multiplying the second
term inside the square brackets, $H_s$ is essentially the same as the
analogous quantity computed in \ref{6k=0}, we therefore do not analyze
the case $\xi=0$ any further and refer the reader to Ref.~\cite{fvv}
for its application to chaotic inflation.
\subsection{Conformal coupling}
\label{conf}
For $\xi=1/6$ one has again a considerable simplification in the
semiclassical equation (\ref{sg}) for $a$,
\be
3\,a\,\left(\dot a^2+k\right)
-{\kappa\over 2}\,\left[{\expec{\hat P_\phi^2}\over a^3}
+a\,\left(a^2\,\omega^2+k\right)\,\expec{\hat\phi^2}\right]
=-{\kappa^2\over12\,a^3}\,\left(\expec{\hat\phi^2\,\hat P_\phi^2}
-\expec{\hat\phi\,\hat P_\phi}^2\right)
\ ,
\label{g6}
\ee
but matter fluctuations do not disappear ($\Delta_\phi\not=0$).
Correspondingly, the Hamiltonian which evolves the matter states is
given by
\be
\hat H_S=
{1\over 2}\,\left[{\hat P_\phi^2\over a^3}
+a\,\left(a^2\,\omega^2+k\right)\,\hat\phi^2
+\left({\expec{\hat\phi\,\hat P_\phi}\over 6\,v\,a^3}
-{\dot a\over a}\right)\,\left(\hat\phi\,\hat P_\phi+
\hat P_\phi\,\hat\phi\right)
-{\hat\phi^2\,\hat P_\phi^2\over 6\,v\,a^3}\right]
\ ,
\label{Hs6}
\ee
and by (\ref{Hs}) (with $\xi=1/6$) when terms of order $v^{-1}$ are
negligible.
In the following we shall assume such an approximation and use the
expressions given in Sections~\ref{matter} and \ref{couple} for the
invariant Fock space with $\xi=1/6$.
\par
We can then study the evolution of the scale factor of the universe
corresponding to a matter content given by $\ket{n}_{1/6}$.
For this we can easily estimate $H_s$ and $H_m$ by employing the
expression for $\sigma_{1/6}$ given by Eq.~(\ref{expan2}).
\subsubsection{Homogeneous mode in flat space}
\label{6k=0}
For $\vec P=k=0$ one has
\be
&&H_m\simeq n\,m_\phi\,
\left(1-{3\,\ell_\phi^2\,\dot a^2\over 8\,a^2}\right)
\label{Hm100}
\\
&&H_s\simeq n\,m_\phi\,
\left(1+{5\,\ell_\phi^2\,\dot a^2\over 8\,a^2}\right)
\label{Hs100}
\ .
\ee
Upon substituting (\ref{Hm100}) into Eq.~(\ref{g6}) we get the trajectory
$a=a_m(\tau)$ as a solution of the master equation
\be
3\,\dot a^2={\ell_p^2\over\ell_\phi}\,
{8\,n\,\ell_\phi\,a^2\over 8\,a^3+n\,\ell_p^2\,\ell_\phi}
\ .
\label{6m}
\ee
Had we used instead (\ref{Hs100}) we would have got a different trajectory
$a=a_s(\tau)$ which solves
\be
3\,\dot a^2={\ell_p^2\over\ell_\phi}\,
{4\,n\,\ell_\phi\,a^2\over 4\,a^3-3\,n\,\ell_p^2\,\ell_\phi}
\ .
\label{6s}
\ee
There is then an obvious difference between $a_m$ and $a_s$, that is the
velocity of the former is always finite for $a\ge0$, while the velocity of
the latter diverges for a finite positive value of $a$ [after the
dilute gas approximation (\ref{dilute}) has broken down].
\par
\begin{figure}
\centerline{
\epsfysize=180pt\epsfbox{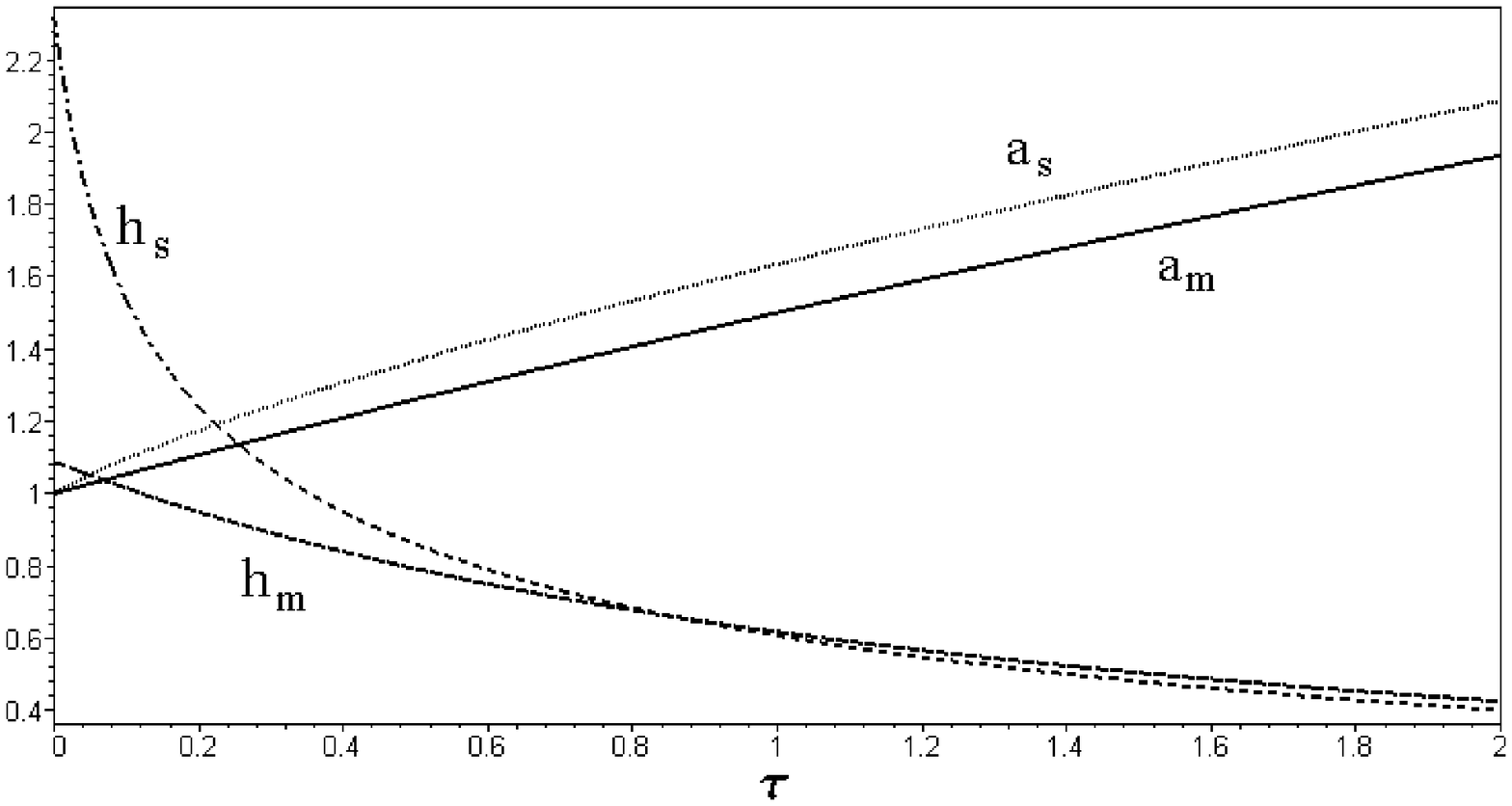}}
\caption{Evolution of the scale factors $a_m$ and $a_s$ for $n=\mu=1$ and
$a(0)=1$; $h_m$ and $h_s$ are the corresponding Hubble coefficients.}
\label{a_ms}
\end{figure}
\begin{figure}
\centerline{
\epsfysize=180pt\epsfbox{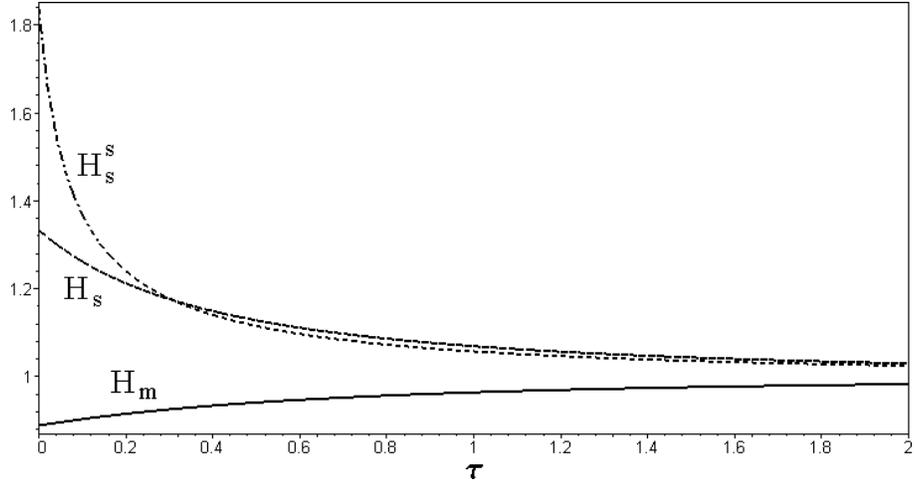}}
\caption{Plot of the three quantities $H_m$, $H_s$ and $H_s^s$ as defined in
the text for the case in Fig.~\ref{a_ms}.}
\label{H_ms}
\end{figure}

In order to show the difference explicitly, we give a first example in
Figs.~\ref{a_ms} and \ref{H_ms}, where we consider a couple of solutions
of the above Eqs.~(\ref{6m}) and (\ref{6s}) with
$\ell_\phi=n\,\ell_p^2\,\ell_\phi=1$ (in natural units $\hbar=\kappa=1$).
Further, we have chosen $a(0)=1$ for both equations so as to avoid the
singularity in $\dot a_s$ and compare two trajectories starting at
the same value.
In Fig.~\ref{a_ms} we show the trajectories $a_m$ and $a_s$ along with the
corresponding Hubble coefficients $h_m=\dot a_m/a_m$ and $h_s=\dot a_s/a_s$
and in Fig.~\ref{H_ms} we plot $H_m[a_m]$, $H_s[a_m]$ and
$H_s^s\equiv H_s[a_s]$.
It is interesting to note that, although the number of invariant quanta,
$n$, remains constant in time for the exact solution $\ket{n}_{1/6}$,
the number of quanta $N\sim H_s$ as computed from the expectation value of
the Hamiltonian $\hat H_S$ decreases and, on the contrary, the ``weight''
$H_m$ of the state $\ket{n}_{1/6}$ increases.
The quantity $H_s^s$ represents what the ``weight'' of the state would be
were the squeezing factor totally absent.
At late times ($\tau\to\infty$) terms proportional to $h_m$ and its
derivatives vanish (adiabatic limit) and the three quantities converge to
the same value.
This is well suited if one aims to study the evolution of the universe
assuming to know its present state [see Eq.~(\ref{in})].
\par
\begin{figure}
\centerline{
\epsfysize=180pt\epsfbox{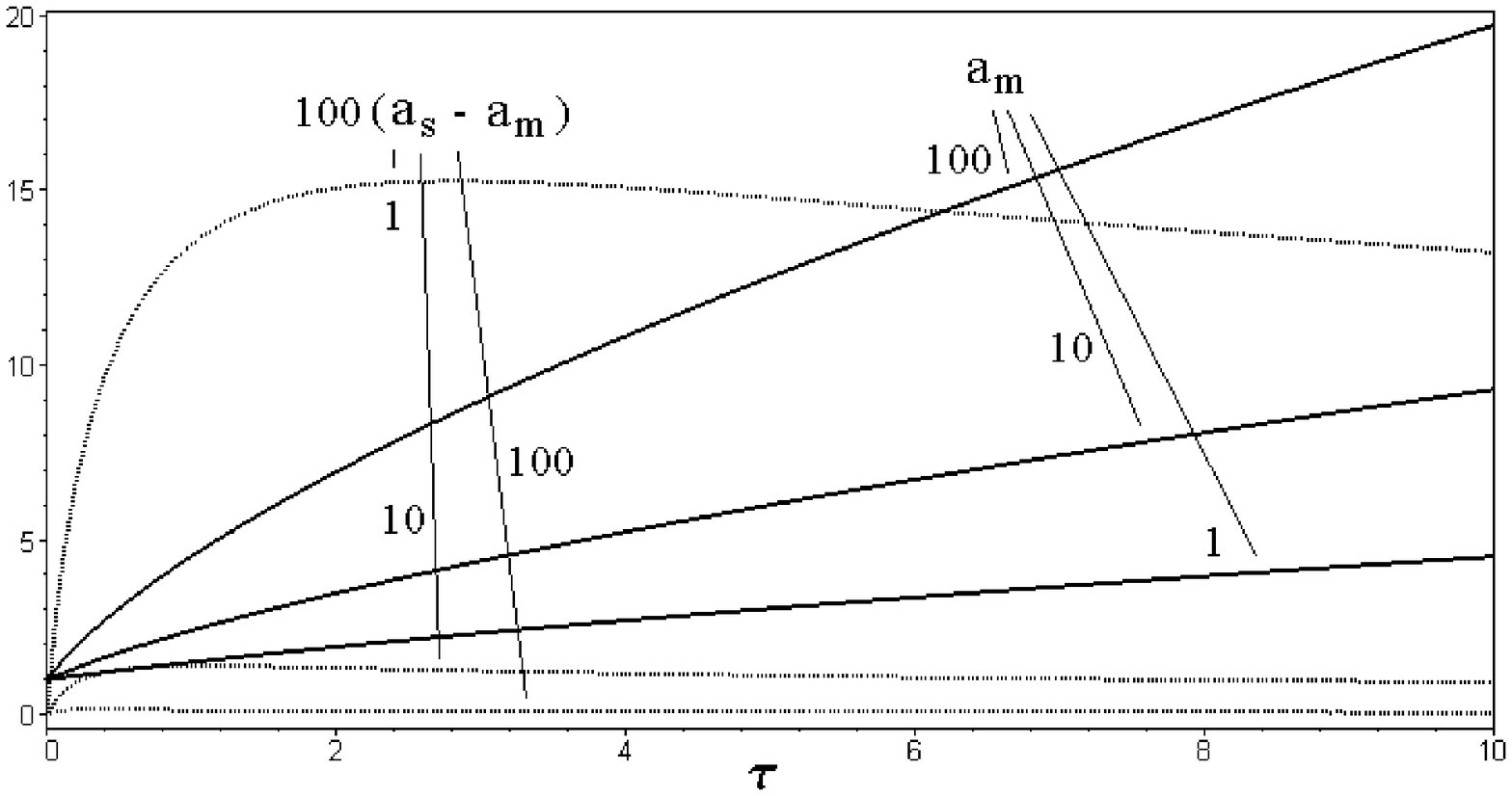}}
\caption{Evolution of the scale factors $a_m$ and $a_s$ for $\mu=1,10,100$.}
\label{mu_ms}
\end{figure}
\begin{figure}
\centerline{
\epsfysize=180pt\epsfbox{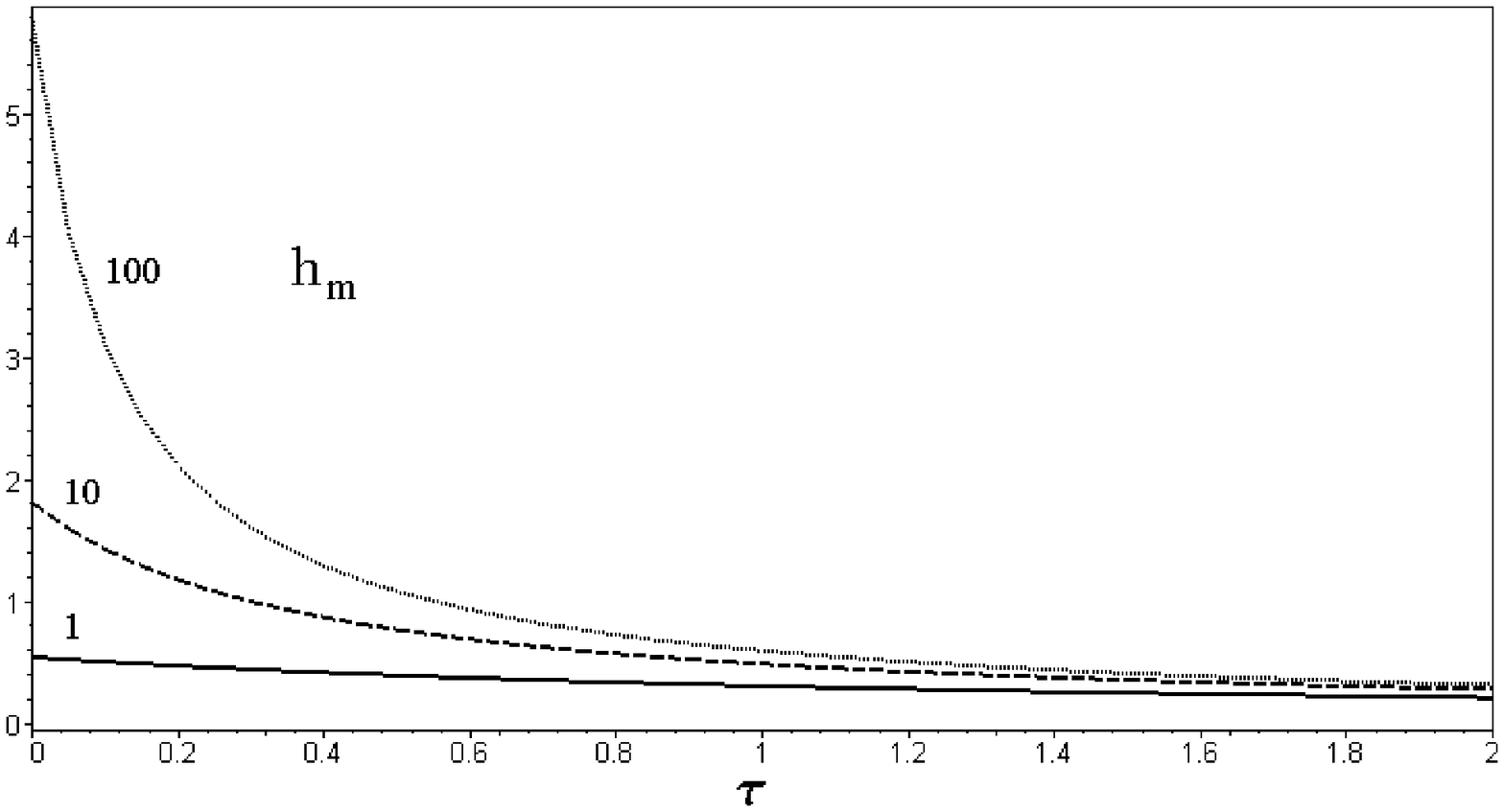}}
\caption{Hubble coefficients $h_m$ for the trajectories $a_m$ in
Fig.~\ref{mu_ms} with $\mu=1,10,100$.}
\label{mu_h}
\end{figure}
\begin{figure}
\centerline{
\epsfysize=180pt\epsfbox{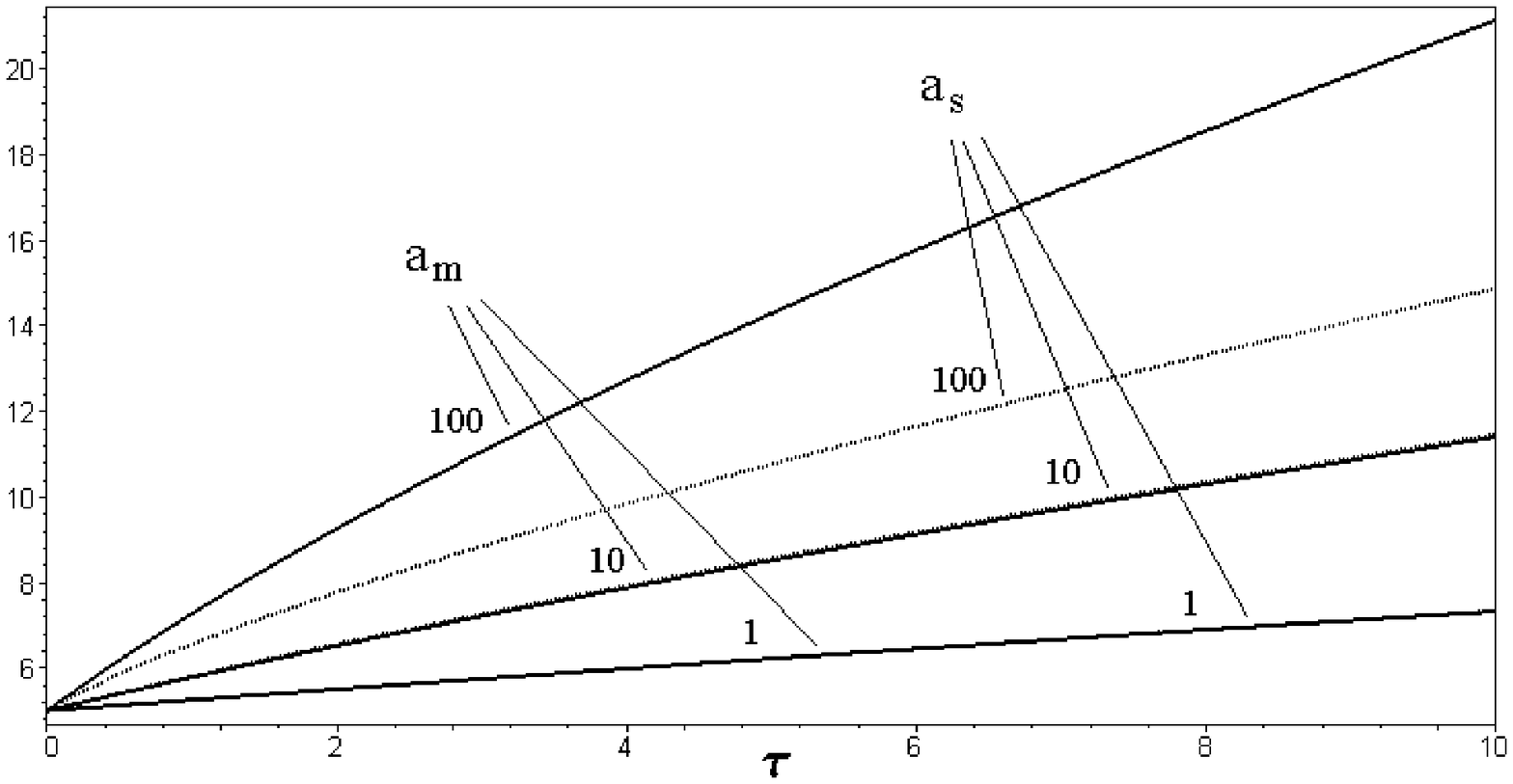}}
\caption{Evolution of the scale factors $a_m$ and $a_s$ for
$n=1,10,100$.}
\label{n_ms}
\end{figure}
\begin{figure}
\centerline{
\epsfysize=180pt\epsfbox{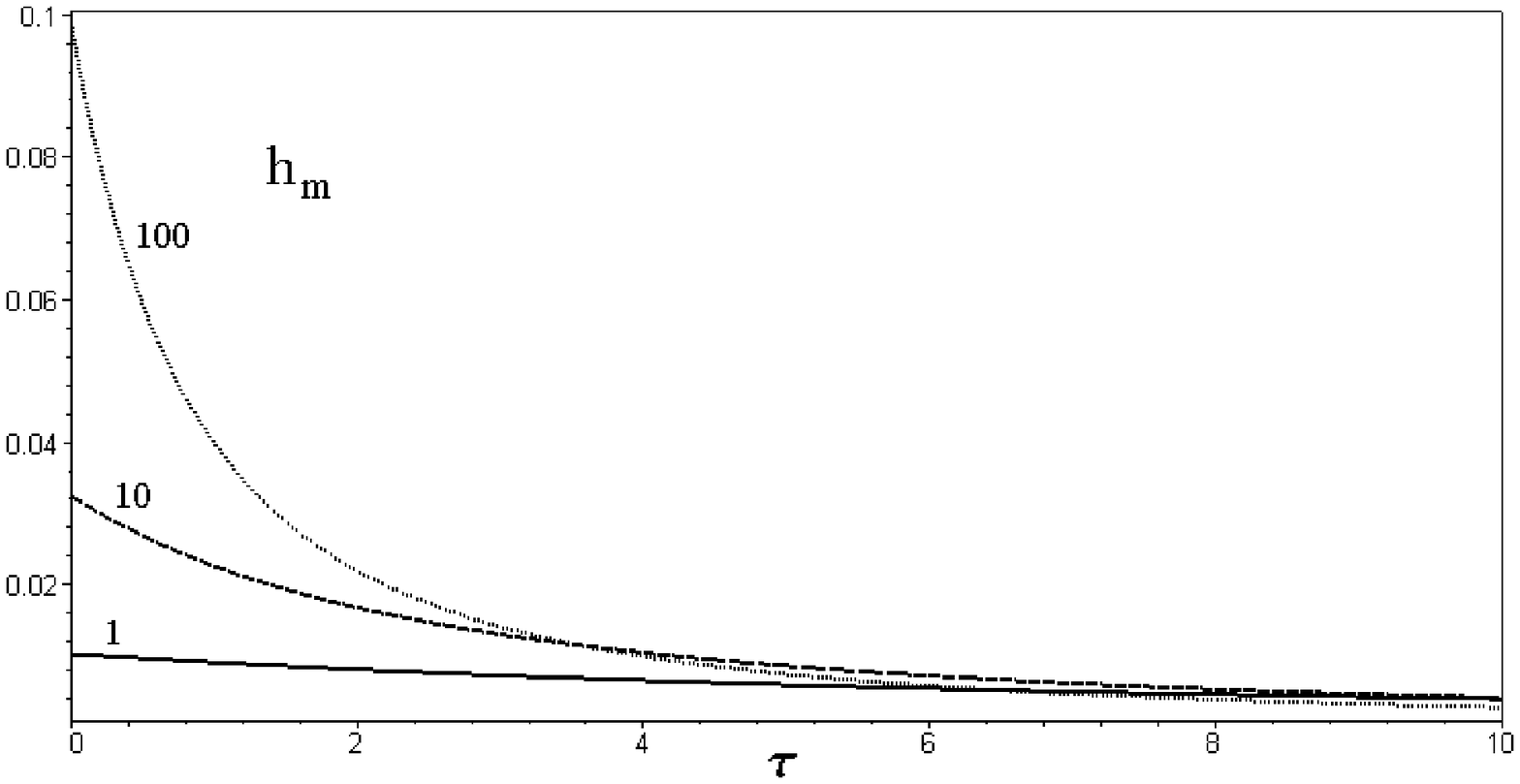}}
\caption{Hubble coefficients $h_m$ for the trajectories $a_m$ in
Fig.~\ref{n_ms} with $n=1,10,100$.}
\label{n_h}
\end{figure}

In Eq.~(\ref{6m}) there are two parameters one can vary, that is $\ell_\phi$
and $n$.
In Fig.~\ref{mu_ms} we show the effect of taking $1/\ell_\phi=\mu=1,10,100$
(with $n\,\ell_p^2\,\ell_\phi=1$) on both $a_m$ and $a_s$ with $a(0)=1$ and
in Fig.~\ref{n_ms} the effect of changing $n=1,10,100$ (with $\mu=1$)
on $a_m$ and $a_s$ with $a(0)=5$.
In Fig.~\ref{mu_h} and \ref{n_h} we plot the corresponding Hubble coefficients
$h_m$.
In particular one can see from Figs.~\ref{n_ms} and \ref{n_h} that both the
scale factor and the Hubble coefficient scale with a positive power of $n$,
as one expected from the fact that $H_m\sim n$.
\par
\begin{figure}
\centerline{
\epsfysize=180pt\epsfbox{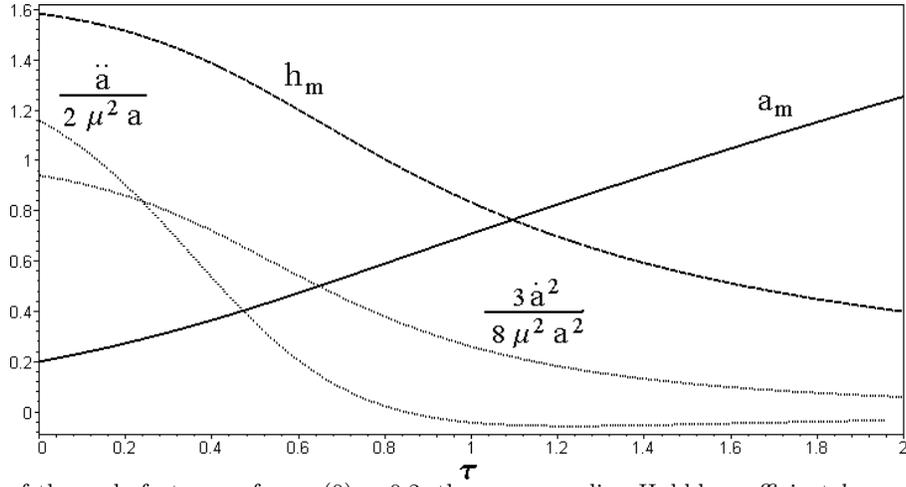}}
\caption{Evolution of the scale factor $a_m$ for $a_m(0)=0.2$, the
corresponding Hubble coefficient $h_m$ and the ratios between velocity and
acceleration of $a_m$ and $\mu$.}
\label{a_m}
\end{figure}
\begin{figure}
\centerline{
\epsfysize=180pt\epsfbox{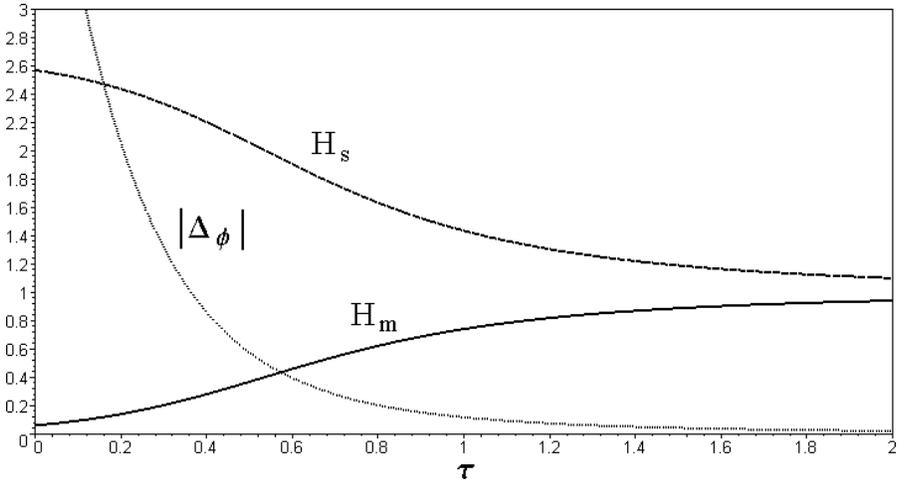}}
\caption{Plot of $H_m$, $H_s$ and $|\Delta_\phi|=-\Delta_\phi$ for the
trajectory in Fig.~\ref{a_m}.}
\label{H_m}
\end{figure}

Since Eq.~(\ref{6m}) does not forbid $a_m(\tau)$ to approach zero, we
also plot in Fig.~\ref{a_m} a trajectory with $\mu=n\,\ell_p^2\,\ell_\phi=1$
which starts at $a_m(0)=0.2$ and the corresponding Hubble coefficient $h_m$.
In the same graph we also show that terms proportional to $\dot a^2/a^2$ and
$(\ddot a/a)\,(\dot a^2/a^2)$ are negligible for the trajectory $a_m$ as is
required by the approximation (\ref{expan2}).
Fig.~\ref{H_m} reproduces the behaviour of $H_m$ and $H_s$.
In particular, it is now apparent that $H_m$ starts out at about zero and
does not diverge for $a_m(0)\to 0$.
However, in this case $n\,\ell_p^2\,\ell_\phi\sim a^3(\tau)$ and the
condition (\ref{dilute}) is violated.
Thus, we have also plotted
\be
-\Delta_\phi\simeq {\ell_p^4\,n^2\over24\,a^3}
\ ,
\label{dphi}
\ee
to compare its relevance with respect to $H_m$.
\par
\begin{figure}
\centerline{
\epsfysize=180pt\epsfbox{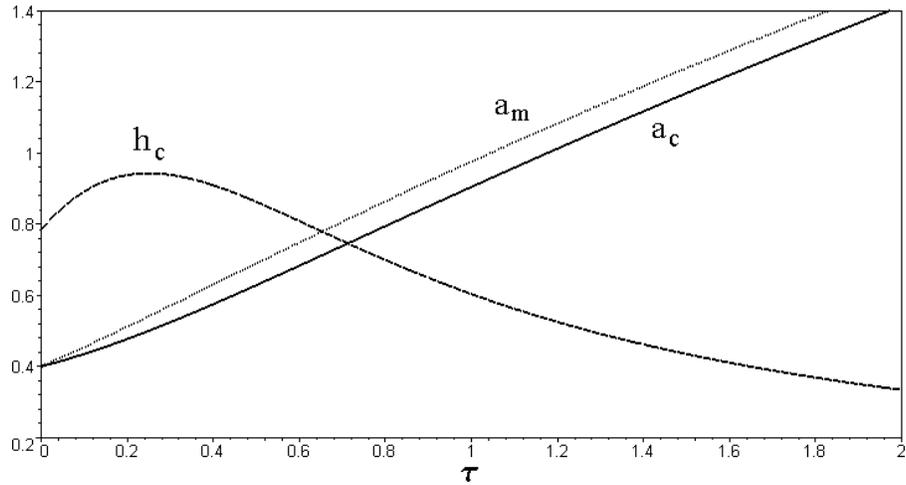}}
\caption{Plot of $a_c$ and $a_m$ with $a(0)=0.4$ and the corresponding
Hubble coefficient $h_c$.}
\label{a_c}
\end{figure}
\begin{figure}
\centerline{
\epsfysize=180pt\epsfbox{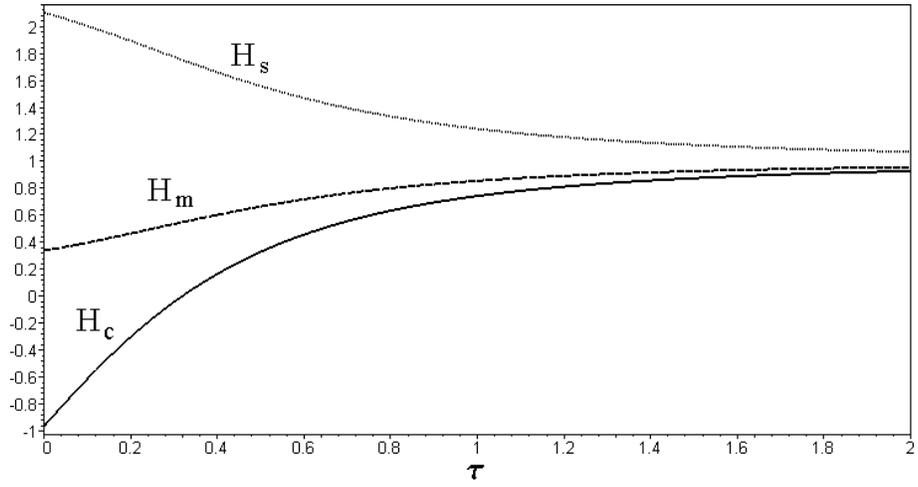}}
\caption{Plot of $H_c$, $H_m$ and $H_s$ for the trajectory $a_c$ in
Fig.~\ref{a_c}.}
\label{H_c}
\end{figure}

Since $\Delta_\phi$ dominates at small $a$, we have computed a corrected
trajectory $a_c=a_c(\tau)$ with $a(0)=0.4$ which we plot in Fig.~\ref{a_c}
together with the corresponding Hubble coefficient $h_c$ and a trajectory
$a_m$ with the same initial condition.
Finally, in Fig.~\ref{H_c} we display the behaviour of
\be
H_c\equiv H_m+\Delta_\phi
\ee
together with $H_m$ and $H_s$ for the
trajectory $a_c$.
\subsubsection{Massless modes}
\label{true6}
This is also a remarkable case, corresponding to what is usually
considered true conformal coupling (because of $\mu=0$).
Indeed, in the approximation (\ref{expan2}) we find that there is
no difference between the gravitational ``weight'' and the expectation
value of the Hamiltonian $\hat H_s$,
\be
H_m\simeq H_s\simeq n\,{\hbar\over a}\,
\left(\vec P\cdot\vec P+k\right)
\ ,
\ee
where $\vec P\cdot\vec P+k$ must be strictly positive [see Eqs.~(\ref{cond}
and (\ref{expan2})].
\par
\begin{figure}
\centerline{
\epsfysize=180pt\epsfbox{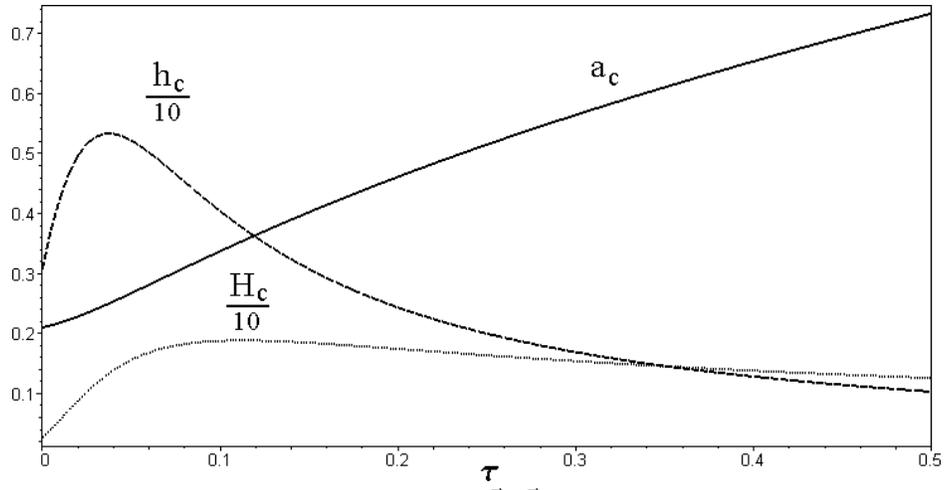}}
\caption{Plot of the trajectory $a_c$ with $\vec P\cdot\vec P=n=1$,
$k=0$ and $a(0)=0.21$.}
\label{a_6}
\end{figure}

Since the term $\Delta_\phi$ in Eq.~(\ref{dphi}) again dominates for small
$a$, the scale factor is determined by the master equation
\be
\dot a^2={n\,\ell_p^2\over 3\,a^2}\,\left(
\vec P\cdot\vec P+k-{n\,\ell_p^2\over 24\,a^2}\right)-k
\ .
\label{eq6}
\ee
In Fig.~\ref{a_6} we plot the solution $a_c=a_c(\tau)$ of Eq.~(\ref{eq6}),
its Hubble coefficient $h_c$ and gravitational ``weight'' $H_c$ for the
mode $\vec P\cdot\vec P=n=1$ in flat space, $k=0$, and $a_c(0)=0.21$.
\subsection{Fermionic coupling}
\begin{figure}
\centerline{
\epsfysize=180pt\epsfbox{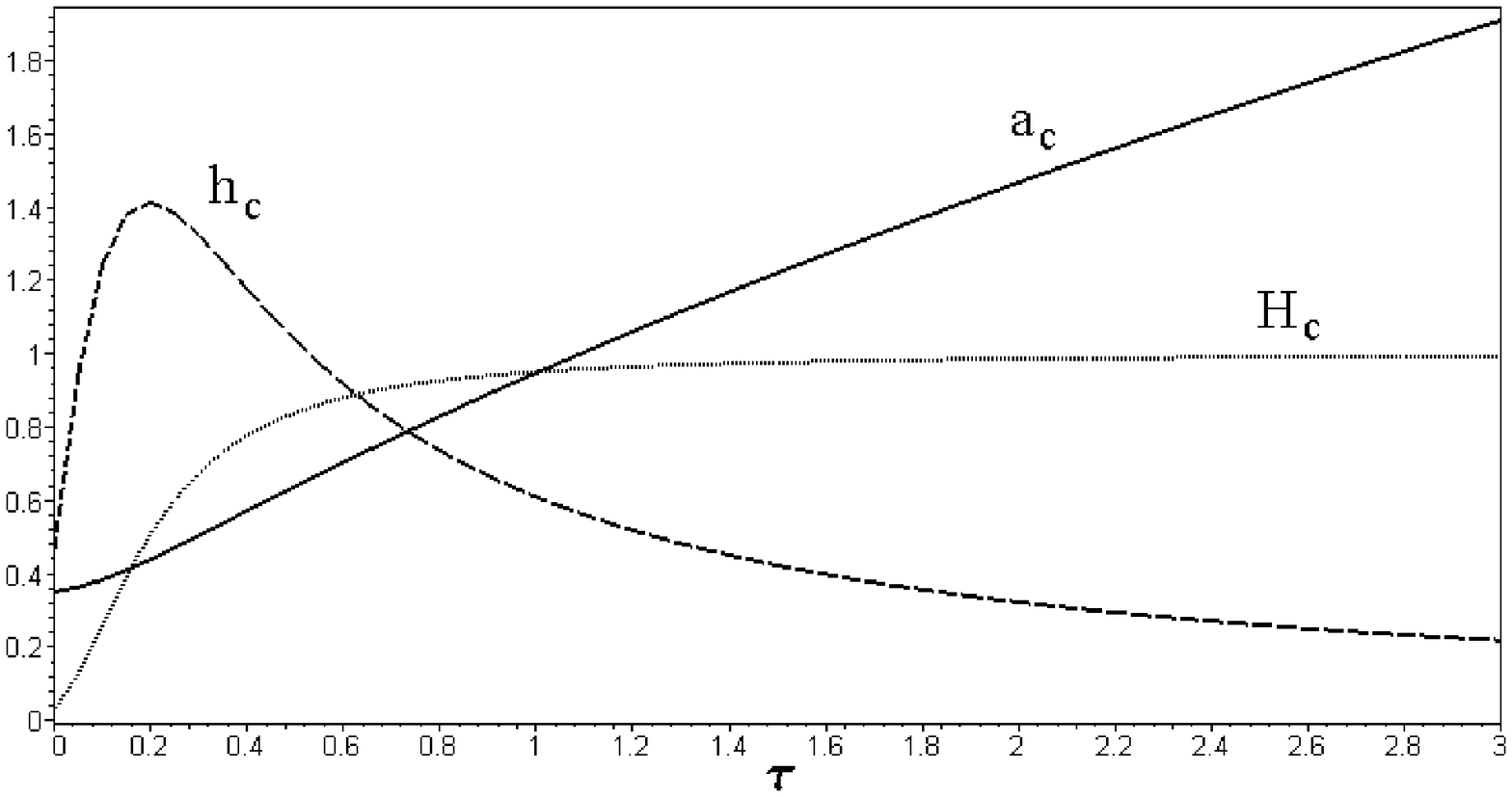}}
\caption{Plot of the trajectory $a_c$ with $a(0)=0.4$, $h_c$ and $H_c$.}
\label{a_4}
\end{figure}

For $\xi=1/4$ and $\vec P=k=0$ one finds that
\be
\sigma_{1/4}={1\over\sqrt{\mu}}
\ee
is an exact solution of Eq.~(\ref{pin}) which holds in the approximation
(\ref{Wv}).
It then follows that
\be
H_m=H_s=n\,m_\phi
\ ,
\ee
which is a constant.
The corresponding master equation for the scale factor, once one includes
$\Delta_\phi$, is given by
\be
3\,\dot a^2=n\,{\ell_p^2\over a}\,\left({1\over\ell_\phi}-
{n\,\ell_p^2\over24\,a^3}\right)
\ .
\ee
We plot the solution $a_c$ for $a(0)=0.4$, in Fig.~\ref{a_4}, together
with the Hubble coefficient $h_c$ and the gravitational ``weight'' $H_c$,
and observe that the qualitative behaviour of these three quantities is
similar to the one of the analogous quantities for the massless
conformally coupled scalar field described in Section~\ref{true6}.
\section{Conclusions}
\label{conc}
In this paper we have analyzed the dynamics of a mode of a real
scalar field non-minimally coupled to the Robertson-Walker metric.
Starting from the Wheeler-DeWitt equation, we have employed the
Born-Oppenheimer approach which
has then led us to a semiclassical picture in which the state of the
scalar field is evolved by a Schr\"odinger equation and the scale factor
of the universe by a semiclassical Hamilton-Jacobi equation.
The main result is that, for generic coupling, the expression for the
gravitational ``weight'' of a matter state is not naively related to
the Hamiltonian operator appearing in the Schr\"odinger equation and
evolves in time differently with respect to the expectation value of
the latter.
Correspondingly, the scale factor of the universe evolves accordingly to
a non-trivial master equation.
\par
By choosing the parameters of the model so as to obtain a Schr\"odinger
equation for which the exact (invariant) Fock space can be constructed
using known methods, we have studied such master equation for the cases
which are mostly treated in the literature, that is the massive minimally
coupled ($\xi=0$) scalar field and both massive and massless scalar
fields with $\xi=1/6$.
Further, we have considered the homogeneous mode of a massive scalar
field with $\xi=1/4$ in flat space whose Klein-Gordon equation
is formally the same as the one satisfied by minimally coupled Dirac
fields and for which the dynamics of the scale factor shows remarkable
qualitative similarities with the case of the massless conformally
coupled ($\xi=1/6$) scalar field.
\par
For $\xi=1/6$ we have explicitly shown that the gravitational ``weight''
$H_m$ of a given massive matter state increases in time, at least during
the early stages of the expansion.
In the spirit of the principle of equivalence, according to which the
gravitational mass of a particle equals its inertial mass, this can be
considered as the signature of real particle production.
In fact, although no local detector has been introduced, one can regard
the scale factor itself as an observable quantity, {\em e.g.}, by means
of measuring the recession of galaxies, and relate the counting of
particles to its evolution.
\par
We also found that $H_s$, the expectation value of the matter Hamiltonian
in the Schr\"odinger equation, generally decreases (or stays constant).
Because of the different behaviours of $H_m$ and $H_s$, one might conclude
that there is a failure of the principle of equivalence, since one would
expect that the energy by which a matter state is evolved in time is the
same that gravitates.
Were this observation proved correct, the massless conformally coupled
scalar field and the homogeneous mode of a massive scalar field
with $\xi=1/4$ in flat space would stand up as very peculiar, since for
them (as for the minimally coupled scalar field) the two quantities are
equal and the equivalence between inertial mass and gravitational mass
would therefore be preserved for the massive case with $\xi=1/4$
(thus suggesting an analogous result for fermions).
\par
However, we point out that, while $H_m$ was shown to be the semiclassical
(time-time component of the) unique covariantly conserved energy-momentum
tensor and has naturally a physical meaning as the energy of the perfect
fluid modeled by the scalar field, $H_s$ cannot be related to any directly
measurable quantities in our treatment.
Hence it is not clear whether $H_s$ carries any physically accessible
information, although the corresponding operator $\hat H_S$ plays a
fundamental role for the dynamics.
In order to enlarge the number of observable quantities and make testable
predictions one might then consider inhomogeneous fluctuations of matter
fields perturbatively on the background determined by the master equations
obtained in this paper and estimate, {\em e.g.}, the effect induced on the
spectrum of the cosmic microwave background radiation.
\par
We wish to conclude by mentioning that further possible extensions of the
present work include a deeper analysis of purely quantum effects, such as
those induced by the superposition of several matter states \cite{super}
or the geometrical phase appearing in Eq.~(\ref{chi_s}) and the r.h.s.s of
Eqs.~(\ref{WdWg}) and (\ref{WdWm}), and different couplings between gravity
and the scalar field, such as those in scalar-tensor theories of gravity
(for a recent review see Ref.~\cite{faraoni}).
All such extensions would affect the evolution of the background and,
eventually, of inhomogeneous fluctuations of the matter fields.
\acknowledgments
We thank G. Venturi for useful discussions.
R.C. thanks V. Frolov for stimulating suggestions.
\appendix
\section{Action for $\Phi_{\vec p}$}
\label{p}
For $\xi=0$ the scalar field potential in the action (\ref{Seh}) is
quadratic in $\Phi$ and different modes in the sum (\ref{sum}) decouple,
while, for $\xi\not=0$, one expects the term $\Phi^2\,R$ induces
interactions via ``graviton exchange''.
In any case, gravity would respond to the sum of all modes,
thus, for the sake of simplicity, we shall consider a scalar field
containing only one mode of fixed wave vector $\vec p$ ,
\be
\Phi={1\over\sqrt{V}}\,\left[\cos\left(\vec p\cdot\vec x\right)\,
\phi_1+\sin\left(\vec p\cdot\vec x\right)\,\phi_2\right]
\ ,
\ee
where $\phi_1\equiv\phi^1_{\vec p}$ and $\phi_2\equiv\phi^2_{\vec p}$.
The scalar product $\vec a\cdot\vec b\equiv \gamma_{ij}\,a^i\,b^j$,
with $\gamma_{ij}=g_{ij}/a^2$ ($i,j=r,\theta,\varphi$) and $\bol g$ as
given in Eq.~(\ref{rw}), is time-independent.
One then finds
\be
\Phi^2={1\over V}\,\left[\cos^2\left(\vec p\cdot\vec x\right)\,
\phi_1^2
+\sin^2\left(\vec p\cdot\vec x\right)\,\phi_2^2
+\sin\left(2\,\vec p\cdot\vec x\right)\,\phi_1\,\phi_2\right]
\ ,
\ee
and an analogous expression for $(\partial\Phi)^2$.
The integration over the spatial volume yields the following constant
coefficients ($\gamma\equiv\det\bol\gamma$)
\be
V_c&\equiv&
\int d^3x\,\sqrt{\gamma}\,\cos^2\left(\vec p\cdot\vec x\right)
\nonumber \\
V_s&\equiv&
\int d^3x\,\sqrt{\gamma}\,\sin^2\left(\vec p\cdot\vec x\right)
\ .
\ee
Since the three-metric $\bol \gamma$ is isotropic, the direction of
$\vec p$ cannot affect the value of the above integrals so that $V_c$
and $V_s$ depend at most on the modulus $\vec p\cdot\vec p$.
Further, homogeneity of $\bol \gamma$ implies that
\be
\int d^3x\,\sqrt{\gamma}\,\sin\left(2\,\vec p\cdot\vec x\right)
=0
\ .
\ee
After recalling that
\be
V\equiv\int d^3x\,\sqrt{\gamma}
\ ,
\ee
and, setting $\phi_2=0$, one then obtains an action for the real part of
$\Phi$,
\be
S_1={1\over 2}\,\int_{t_i}^{t_f} N\,dt\,a^3\,\left[\left(
{V_c\over V}\,{\dot\phi_1^2\over N^2}-\omega_1^2\,\phi_1^2\right)
+\left({V\over\kappa}-\xi\,{V_c\over V}\,\phi_1^2\right)\,R\right]
\ ,
\ee
with
\be
\omega_1^2\equiv
{V_s\over V}\,{\vec p\cdot\vec p\over a^2}+{V_c\over V}\,\mu^2
\ ,
\ee
and, setting $\phi_1=0$, an action for the imaginary part,
\be
S_2={1\over 2}\,\int_{t_i}^{t_f} N\,dt\,a^3\,\left[\left(
{V_s\over V}\,{\dot\phi_2^2\over N^2}-\omega_2^2\,\phi_2^2\right)
+\left({V\over\kappa}-\xi\,{V_s\over V}\,\phi_2^2\right)\,R
\right]
\ ,
\ee
with
\be
\omega_2^2\equiv
{V_c\over V}\,{\vec p\cdot\vec p\over a^2}+{V_s\over V}\,\mu^2
\ .
\ee
\par
For the homogeneous mode, $\vec p=0$, one has $V_c=V$ and $V_s=0$, so that
$S_2$ vanishes and $S_1$ coincides with the expression in Eq.~(\ref{Sp})
with $\phi\equiv\phi_1$ and $\omega=\mu$.
\par
For $\vec p\not=0$, both $V_c$ and $V_s$ are strictly positive and one can
rescale the fields according to
\be
\phi\equiv\sqrt{{V_c\over V}}\,\phi_1
\ \ \ \ \left(\hbox{or}\ \
\phi\equiv\sqrt{{V_s\over V}}\,\phi_2\right)
\ ,
\ee
and correspondingly define an ``effective'' wave vector
\be
\vec P\equiv\sqrt{{V_s\over V_c}}\,\vec p
\ \ \ \ \left(\hbox{or}\ \
\vec P\equiv\sqrt{{V_c\over V_s}}\,\vec p\right)
\ ,
\ee
so that the action $S_1$ ($S_2$) for the real (imaginary) part $\phi$ is
again equal to the expression in Eq.~(\ref{Sp}) with
\be
\omega^2={\vec P\cdot\vec P\over a^2}+\mu^2
\ .
\label{freq}
\ee
This shows that the action (\ref{Sp}) can be used to describe the dynamics
of (the real or imaginary part of) each mode of the real scalar field.
\section{Boundary terms}
\label{border}
The procedure which leads to the action (\ref{S}) from the one in
Eq.~(\ref{Sp}) is the analogue (for generic $\xi$) of what is done
in general relativity when one defines the Einstein-Hilbert action
as \cite{mtw}
\be
S_{EH}={1\over2\,\kappa}\,\int_{\cal M} d^4x\,\sqrt{-g}\,R
-{1\over\kappa}\,\int_{\partial{\cal M}} d^3x\,K
\ ,
\ee
where $K$ is the extrinsic curvature of the border $\partial{\cal M}$
of the space-time manifold ${\cal M}$.
In the above, the surface integral evaluated on $\partial{\cal M}$
precisely cancels all the troublesome terms inside the volume
contribution (including first time derivatives of the lapse
function and second time derivatives of the three-metric).
\par
For the Robertson-Walker metric (\ref{rw}) one has that $K$ vanishes
at the time-like border $r=r_k$ and the only contribution to the surface
integral comes from the hypersurfaces $t=t_i$ and $t=t_f$,
\be
{1\over\kappa}\,\int_{\partial{\cal M}} d^3x\,K
&=&3\,{V\over\kappa}\,\left[a^2\,{\dot a\over N}\right]^{t_f}_{t_i}
\nonumber \\
&=&{1\over 2}\,\int_{t_i}^{t_f} dt\,{d\over dt}\,\left[6\,v\,
a^2\,{\dot a\over N}\right]
\ .
\ee
It therefore appears natural, for $\xi\not=0$, to generalize the
standard prescription to
\be
{1\over\kappa}\,\int_{\partial{\cal M}} d^3x\,K
\to{1\over 2}\,\int_{t_i}^{t_f} dt\,{d\over dt}\,\left[6\,
\left(v-\xi\,\phi^2\right)\,a^2\,{\dot a\over N}\right]
\ ,
\ee
in order to eliminate unwanted terms from the action (\ref{Sp})
and obtain the form (\ref{S}).
\par
Of course one could also consider other ways of proceeding and,
{\em e.g.}, allow for terms containing $\ddot a$.
However, the requirement that time-reparameterization remains an
invariance of the system ($P_N=0$) seems to favor the above
procedure.
\newpage
\newpage
\newpage
\newpage

\begin{references}
\bibitem{full}
We remark that the full scalar field $\Phi$ couples to gravity for
$\xi\not=0$.
Alternatively one might want to couple $R$ only to quantum
fluctuations by setting $\Phi=\Phi_c+\delta\Phi$ and replacing the
factor $1/\kappa-\xi\,\Phi^2$ which multiplies $R$
with $1/\kappa-\xi\,\delta\Phi^2$, but we shall not attempt at this
in the present paper.
\bibitem{miss}
In this paper we shall refer to the case with $\xi=1/6$ as to the
conformally coupled scalar field, although, strictly speaking, such
denomination also requires $\mu=0$ \cite{birrell}.
\bibitem{diago}
One could try to define a new canonical variable $\alpha$ such that
its conjugate momentum is $\Pi_\alpha=P_a-6\,\xi\,\phi\,P_\phi/a$
and diagonalizes the kinetic term in Eq.~(\ref{H}).
However, we observe that the explicit form of $\alpha$, as determined
by the condition
$\Pi_\alpha\,\dot\alpha+ P_\phi\,\dot\phi=
P_a\,\dot a+ P_\phi\,\dot\phi+\dot F$,
where $F$ is an arbitrary function, would shift the Hamiltonian,
$H\to H-\dot F$, and necessarily mix the gravitational and matter
degrees of freedom (for $\xi\not=0$), to wit $\alpha=\alpha(a,\phi;\xi)$.
The latter fact places serious questions on the definition of the
semiclassical limit for gravity as described in the Introduction,
since one cannot disentangle matter from gravity after such a
canonical transformation has been performed.
\bibitem{obs}
One could avoid most of such formal problems by assuming the
(semiclassical) variable $a$ is eventually the only physical observable
in the system.
Since $a$ determines the relative positions of freely falling (comoving)
points in the Robertson-Walker manifold and space-time points in general
relativity are defined only when a material reference frame is introduced
\cite{mtw,hk},
this also requires the existence of some (semi)classical matter whose
``weight'' we shall identify in Eq.~(\ref{Tf}) with the quantity $H_m$
given in Eq.~(\ref{Hmn}).
\bibitem{cl}
More precisely, there is only one state $\ket{\bar n}$ such that
$\delta a_{\bar n}\not=0$ and this gives the classical evolution with
$\dot a\,\delta t\sim\delta a_{\bar n}$.
\bibitem{super}
One can also consider a generic superposition of invariant eigenstates
$\chi_s=\sum\,c_n\,\ket{n,\tau}_s$, in which case one usually obtains
oscillating contributions to $H_s$ (and $H_m$) \cite{cfv,fvv}.
\end{references}
\end{document}